\documentstyle[sprocl,twoside]{article}
\newcommand{\be}{\begin{equation}}
\newcommand{\ee}{\end{equation}}
\newcommand{\bee}{\begin{eqnarray}}
\newcommand{\eee}{\end{eqnarray}}

\newcommand{\ga}{\alpha}
\newcommand{\gge}{\epsilon}
\newcommand{\gb}{\beta}
\newcommand{\gga}{\gamma}

\newcommand{\gd}{\delta}

\newcommand{\gep}{\epsilon}
\newcommand{\gvep}{\varepsilon}
\newcommand{\gs}{\sigma}
\newcommand{\go}{\omega}

\newcommand{\q}{\,,\qquad}
\newcommand{\M}{{\cal M}}
\newcommand{\X}{{\cal X}}
\newcommand{\ls}{\!\!\!\!\!\!}

\newcommand{\da}{{\dot{a}}}
\newcommand{\db}{{\dot{b}}}
\newcommand{\dc}{{\dot{c}}}
\newcommand{\dd}{{\dot{d}}}

\newcommand{\bZ}{\overline{Z}}

\newcommand{\nn}{\nonumber}
\newcommand{\half}{\frac{1}{2}}

\newcommand{\p}{\partial}

\newcommand{\D}{{\cal D}}

\newcommand{\f}{\frac}

\newcommand{\C}{{\cal C}}

\newcommand{\rvac}{|0\rangle}
\newcommand{\lvac}{\langle 0|}

\begin{document}
\sloppy

\title
{RELATIVITY, CAUSALITY, LOCALITY,
QUANTIZATION\\ AND DUALITY\\ IN THE $Sp(2M)$ INVARIANT
GENERALIZED SPACE-TIME}

\author{M.A.Vasiliev}

\address
{I.E.Tamm Department of Theoretical Physics, Lebedev Physical
Institute,\\
Leninsky prospect 53, 119991, Moscow, Russia}

\maketitle\abstracts
{We analyze properties of the $Sp(2M)$ conformally invariant field
equations in the recently proposed generalized  $\half M(M+1)-$
dimensional space-time $\M_M$ with matrix coordinates.  It is shown
that classical solutions of these field equations define a causal
structure in $\M_M$ and admit a well-defined decomposition into
positive and negative frequency solutions that allows consistent
quantization in a positive definite Hilbert space.  The effect of
constraints on the localizability of fields in the generalized
space-time is analyzed. Usual $d-$dimensional Minkowski space-time is
identified with the subspace of the matrix space $\M_M$ that allows
true localization of the dynamical fields. Minkowski coordinates are
argued to be associated with some Clifford algebra in the matrix space
$\M_M$. The dynamics of a conformal scalar and spinor in $\M_2$ and
$\M_4$ is shown to be equivalent, respectively, to the usual conformal
field dynamics of a scalar and spinor in the $3d$ Minkowski space-time
and the dynamics of massless fields of all spins in the $4d$ Minkowski
space-time. An extension of the electro-magnetic duality transformations
to all spins is identified with a particular generalized Lorentz
transformation in $\M_4$. The $M=8$ case is shown to correspond to a
$6d$ chiral higher spin theory. The cases of $M=16$ (d=10) and
$M=32$ (d=11) are discussed briefly.}

\newpage

\tableofcontents

\newpage

\section*{To the memory of Mikhail Marinov}

During several years I had an opportunity
to appreciate attractiveness and power
of the personality of Misha Marinov. Participating in the same
seminars at Lebedev Physical Institute and Moscow State
University we discussed physics many times. I was particularly
happy to be able to discuss with Misha some questions related
to the theory of higher spin gauge fields, such as star product
algebras, symbols of operators and others. These discussions gave
me a lot. Misha's attitude to a scientific question always
depended on to which extend he believed the topic
belonged to a true physics area. Of course, every physicist
has his own feeling of ``true physics". An ``invariant" principal
part Misha insisted on was that there always must be  a particular
physical problem behind any formal
manipulations. That was certainly a good school.  I hope
that this contribution  fits the high standard of Marinov.

After Misha went to the Land of Fathers one obvious and not at
all surprising change I noticed during our short
but full of discussions meeting at a conference in 1996
was how much he appreciated and enjoyed to be an
independent Man, the feeling he was not allowed to have while living in
the Soviet Union. It is too unfair that Misha was not given longer time
to live full life and too unfortunate that no one can any
longer discuss physics to Mikhail Marinov.

\newpage

\section{Introduction}
\label{Introduction}

In the recent paper \cite{BHS} the system of conformally invariant
equations of motion for $4d$ massless fields of all spins
(for more details on the theory of $4d$ higher spin gauge fields
we refer the reader to \cite{Gol}) was shown to
exhibit generalized conformal symmetry \footnote{We use
notation $Sp(2M)$  for
the noncompact real group $Sp(2M,{\bf R})$ constituted
by $2M\times 2M$ real matrices that leave invariant a non-degenerate
real $2M\times 2M$ antisymmetric bilinear form.} $Sp(8)$
and was argued  to be equivalent to a system of equations
for scalar and spinor in the generalized ten-dimensional space-time.
An extension of these equations to the
generalized $\half M(M+1)-$ dimensional space-time
with arbitrary even $M$ was proposed in the same
reference \cite{BHS} \footnote{Note that $M$ is required to be even by
the $AdS$ version of the model discussed in \cite{BHS}.}.
Based on the idea of Bogolyubov transform
duality \cite{3d} between classical and quantum pictures,
the proposed equations in $\M_M$ were argued in
\cite{BHS} to admit a consistent quantization in a positive definite
Hilbert space. The aim of this paper is to reconsider
the $Sp(2M)$ invariant equations in $\M_M$ from the perspective
of the standard  field-theoretical approach. We will show that these
equations  define a
causal structure of the generalized space-time and admit consistent
quantization with positive and negative frequency solutions
giving rise to the creation and annihilation operators in a
positive definite Hilbert space. Usual Minkowski space-time will
be identified with some $d\leq M+1-$dimensional
submanifold of the generalized space-time
that admits localization of fields. {}From this perspective, the
usual space-time can be thought of as a visualization of the
generalized space-time. Remarkably,  the
space-time submanifolds that admit localization turn out to be
related to certain Clifford algebras which, in turn, give rise to
the usual Minkowski geometry.

The formulations of $Sp(2M)$ invariant systems
in terms of the generalized space-time $\M_M$
and usual space-time are equivalent and
complementary. The description in terms of $\M_M$ provides clear
geometric origin for the $Sp(2M)$ generalized conformal symmetry.
In particular it provides a geometric interpretation of the
electromagnetic duality transformations as particular generalized
Lorentz transformations. However, to define true local
fields, one has to resolve some constraints.
The description in terms of the Minkowski space-time,
that solves the latter problem, makes some of the symmetries
not manifest.

The study of the dynamical equations in $\M_M$ is
likely to be of key importance for the analysis of dynamical systems
that exhibit $Sp(2M)$ symmetries. In particular, the formulation
in terms of generalized space-time $\M_M$ is expected to be useful
for the investigation of the M-theory through the
algebras $sp(32)$, $sp(64)$ and their superextensions \cite{Malg,Bars}.
Various ideas on a possible structure of alternative to
Minkowski space-times have appeared both in the field-theoretical
\cite{H,F,AS} and world particle dynamics
\cite{PD} contexts. In particular, relevance of
a $Sp(8)$ invariant 10-dimensional space-time for the
description of the massless fields of all spins in four dimensions
was emphasized by Fronsdal in \cite{F} where also a realization
of $Sp(2M)$ in the $\half M(M+1)-$ dimensional manifold formed by
isotropic $M-$forms was given, which construction is
closely related to the one discussed in this paper.  To the
best of our knowledge, the dynamical equations we study in this paper,
that allow for physical interpretation of the generalized space-time,
were first suggested in \cite{BHS}.

The generalized  flat $\half M (M+1)-$dimensional
space-time $\M_M$ is described by the matrix coordinates
$X^{\ga\gb}=X^{\gb\ga}$ ($\ga ,\gb = 1\ldots M$).
The $Sp(2M)$  generalized conformal symmetry transformations
in $\M_M$ are realized by the vector fields \cite{BHS}
\be
\label{PS}
 P_{\ga\gb} =-i \f{\p}{\p X^{\ga\gb}}\,,
\ee
\be
L_\ga{}^\gb = 2i
 X^{\gb\gga} \f{\p}{\p X^{\ga \gga}}\,,
\ee
\bee
\label{PK}
K^{\ga\gb}=  -i X^{\ga\gga}
X^{\gb{\eta}}\f{\p}{\p X^{\gga{\eta}}}\,.
\eee
The (nonzero) $sp(2M)$ commutation relations are
\be
\label{comLL}
[L_\ga{}^\gb \,, L_\gga{}^\gd] = i\left( \gd^\gd_\ga L_\gga{}^\gb -
\gd^\gb_\gga L_\ga{}^\gd \right )\,,
\ee
\be
\label{comL}
[L_\ga{}^\gb \,, P_{\gga\gd}] = -i \left (\gd^\gb_\gga P_{\ga\gd}
+\gd^\gb_\gd P_{\ga\gga}\right )\,,\qquad
[L_\ga{}^\gb \,, K^{\gga\gd}] =i \left (\gd^\gga_\ga K^{\gb\gd}+
\gd_\ga^\gd K^{\gb\gga}\right )\,,\qquad
\ee
\be
\label{comPK}
[P_{\ga\gb} \,, K^{\gga\gd}] = \f{i}{4}
\Big (\gd_\gb^\gga L_\ga{}^\gd + \gd_\ga^\gga L_\gb{}^\gd +
\gd_\ga^\gd L_\gb{}^\gga + \gd_\gb^\gd L_\ga{}^\gga \Big ) \,.
\ee

Here $P_{\ga\gb}$ and $K^{\ga\gb}$ are generators of the
generalized translations and special conformal transformations.
The $gl_M$ algebra spanned by $L_\ga{}^\gb $
decomposes into the central subalgebra associated with the
generalized dilatation generator
\be
D= L_\ga{}^\ga \nn
\ee
 and the  $sl_M$ generalized  Lorentz generators
\be
l_\ga{}^\gb = L_\ga{}^\gb -\f{1}{M}\delta^\gb_\ga  D\,.\nn
\ee

The infinitesimal transformations generated by the vector fields
(\ref{PS})-(\ref{PK}) can be integrated to the finite group
transformations
\be
\label{Tgl}
X^{\ga\gb}\rightarrow \tilde{X}^{\ga\gb}=X^{\ga\gb} +a^{\ga\gb}
\ee
for generalized translations,
\be
\label{Lgl}
X^{\ga\gb}\rightarrow \tilde{X}^{\ga\gb} =a^{\ga}{}_\gga a^{\gb}{}_\gd
X^{\gga\gd}
\ee
for generalized Lorentz $SL_M$ transformations ($ det| a^{\ga}{}_\gga |=1$)
or dilatations ($a^{\ga}{}_\gga = { \kappa} \delta^\ga{}_\gga$,
${ \kappa} \neq 0$), and
\be
\label{Kgl}
X^{\ga\gb}\rightarrow \tilde{X}^{\ga\gb}=(X_{\ga\gb} +a_{\ga\gb})^{-1}\,
\ee
for generalized special conformal transformations
where $X_{\ga\gb}$ is the inverse to $X^{\ga\gb}$
\be
X_{\ga\gga}X^{\gga\gb}  =\delta_\ga^\gb\,.\nn
\ee
Like in the standard conformal transformations the full
generalized conformal group contains inversion $R$
\be
\label{inv}
R(X^{\ga\gb}) = X_{\ga\gb}\,,
\ee
which is involutive
\be
R\circ R = Id\,. \nn
\ee

A generalized
special conformal transformation $S(a_{\ga\gb})$ can be
represented as a combination of two inversions and some translation
\be
S(a_{\ga\gb})= R\circ T(a_{\ga\gb})\circ R \,.\nn
\ee
It is possible to define the action of the $Sp(2M)$ transformations on
a generalized tensor field
$\phi^{\ga_1 \ldots \ga_n}{}_{\gb_1 \ldots \gb_m} (X)$ as follows.
A finite translation $T(a)$ with the parameter $a^{\ga\gb}$ is defined as
usual
\be
T(a)\Big (\phi^{\ga_1 \ldots \ga_n}{}_{\gb_1 \ldots \gb_m} (X) \Big )=
\phi^{\ga_1 \ldots \ga_n}{}_{\gb_1 \ldots \gb_m} (X+a)\,. \nn
\ee
A finite $GL_M$ transformation with the parameter $a^\ga{}_\gb$ that contains
generalized Lorentz transformations ($det\Big |a \Big |=1$) and dilatations
($a^\ga{}_\gb\sim \gd^\ga_\gb$) is
\bee
\label{lorb}
&{}&\ls\ls\ls G(a)\Big (\phi^{\ga_1 \ldots \ga_n}{}_{\gb_1 \ldots \gb_m}
(X^{\nu\mu})\Big ) =\nn\\ &{}& \ls\ls\ls\Big ( det\Big |a \Big | \Big )^\Delta
a^{-1\,\ga_1}{}_{\gga_1} \ldots a^{-1 \ga_n}{}_{\gga_n}a^{\gd_1}{}_{\gb_1}
\ldots a^{\gd_m}{}_{\gb_m} \phi^{\gga_1 \ldots \gga_n}{}_{\gd_1 \ldots \gd_m}
(a^\nu{}_\eta a^\mu{}_\gs X^{\eta\gs})\,,
\eee
where the parameter $\Delta$ is the generalized conformal weight of the
tensor field $\phi$. A finite generalized special conformal transformation
with the parameter $a_{\ga\gb}$ is
\bee
&{}&\ls\ls\ls S(a)\Big (\phi^{\ga_1 \ldots \ga_n}{}_{\gb_1 \ldots \gb_m}
(X^{\nu\mu})\Big ) =\nn\\&{}&\ls\ls\ls \Big ( det\Big |Q \Big | \Big )^\Delta
Q^{-1\,\ga_1}{}_{\gga_1} \ldots Q^{-1 \ga_n}{}_{\gga_n}Q^{\gd_1}{}_{\gb_1}
\ldots Q^{\gd_m}{}_{\gb_m} \phi^{\gga_1 \ldots \gga_n}{}_{\gd_1 \ldots \gd_m}
(Q^\nu{}_\eta  X^{\eta\mu})\,,
\eee
where
\be
\label{Q}
Q^{-1\,\ga}{}_\gb =\gd^\ga_\gb +X^{\ga\gga} a_{\gga\gb}           \,.
\ee
Finally, the inversion (\ref{inv})
interchanges upper and lower indices according to
the rule
\bee
&{}&\ls \ls R\Big (\phi^{\ga_1 \ldots \ga_n}{}_{\gb_1 \ldots \gb_m} (X)\Big )
=
\tilde{\phi}_{\ga_1 \ldots \ga_n}{}^{\gb_1 \ldots \gb_m} (X^{-1})\Big )
=\nn\\
&{}&\ls\ls\ls\Big ( det\Big |X \Big | \Big )^{-\Delta} X^{-1}{}_{\ga_1\gga_1}
\ldots X^{-1}{}_{\ga_n \gga_n}X^{\gd_1\gb_1} \ldots X^{\gd_m \gb_m}
\phi^{\gga_1 \ldots \gga_n}{}_{\gd_1 \ldots \gd_m} (  X^{-1})\,.
\eee

The singularities of  special conformal transformations
are localized where the matrix $Q^{-1 \ga}{}_{\gb}$ (\ref{Q})
degenerates. To define a globally defined action of the conformal
group the generalized space-time $\M_M$ has to be compactified to
$\C\M_M$ by adding the ``infinity" strata associated with the
degenerate matrices $X^{\ga\gb}$.
(The generalized conformal infinity is a stratified
manifold  because the equations that single out the space of degenerate
real matrices of a given rank  are singular). Compactified matrix spaces
were defined e.g. in \cite{H,F}.
A simple coset space realization
of $\C\M_M$ is given in section \ref{CM}.

The equations of motion proposed in \cite{BHS} read
\be
\label{oscal} \Big (
\f{\p^2}{\p X^{\ga\gb} \p X^{\gga\gd}} - \f{\p^2}{\p X^{\ga\gga} \p
X^{\gb\gd}}\Big ) b(X) =0
\ee
for a scalar field $b(X)$ and
\be
\label{ofer} \f{\p}{\p X^{\ga\gb}} f_\gga(X)       -
\f{\p}{\p X^{\ga\gga}} f_\gb(X)      =0
\ee
for a svector field $f_\gb(X)$.
(We use the name ``svector" (symplectic vector)
to distinguish $f_\gb(X)$
from  vectors of the usual Lorentz algebra $o(d-1,1)$.
Note that svector fields will be shown to obey the Fermi statistics).
For $M=2$, because antisymmetrization of any two-component indices
$\ga$ and $\gb$ is equivalent to their contraction with
the $2\times 2 $ symplectic form $\gep^{\ga\gb}$, (\ref{oscal})
and (\ref{ofer}) coincide with the $3d$ massless Klein-Gordon and Dirac
equations, respectively. For $M=4$,  the equations
(\ref{oscal}) and (\ref{ofer}) in the generalized ten-dimensional
space-time $\M_4$ were argued in
\cite{BHS} to encode the infinite set of the usual
$4d$ equations of motion for massless  fields of all spins.

The equations (\ref{oscal}) and (\ref{ofer})
are invariant under the $Sp(2M)$  generalized
conformal symmetry transformations provided that both $b(X)$ and
$f_\ga (X)$ have conformal weight $\Delta = \half$.
The invariance under generalized translations and Lorentz transformations
is obvious. To prove the full invariance it is enough to check
that the equations (\ref{oscal}) and (\ref{ofer}) are invariant under
generalized dilatations and inversions.
The infinitesimal transformations are \cite{BHS}
\bee
\label{db}
\delta b(X) &=&\Big ( \gvep^{\ga\gb} \f{\p}{\p X^{\ga\gb}}
+\half\gvep^{\ga}{}_{\ga}+2\gvep^{\ga}{}_{\gb} X^{\gb\gga}
\f{\p}{\p X^{\ga \gga}}\nn\\&-&
\gvep_{\ga\gb}\Big [\half X^{\ga\gb} + X^{\ga\gga}
X^{\gb{\eta}}\f{\p}{\p X^{\gga{\eta}}}\Big ] \Big )b(X)\,,
\eee
\bee
\label{df}
\delta f_\gga(X)\ls\, &=&\ls\,\Big ( \gvep^{\ga\gb} \f{\p}{\p X^{\ga\gb}}
+\half\gvep^{\ga}{}_{\ga}+2\gvep^{\ga}{}_{\gb} X^{\gb\eta}
\f{\p}{\p X^{\ga \eta}}\nn\\&-&
\gvep_{\ga\gb}\Big [ \half X^{\ga\gb} + X^{\ga\delta}
X^{\gb{\eta}}\f{\p}{\p X^{\delta{\eta}}}\Big ] \Big )f_\gga(X)
+\left (\gvep^{\gb}{}_\gga - \gvep_{\eta\gga}X^{\eta\gb} \right )f_\gb ,
\eee
where $\gvep^{\ga\gb}$, $\gvep^\ga{}_\gb$ and $\gvep_{\ga\gb}$ are,
respectively, $X-$independent
parameters of generalized translations, Lorentz transformations along with
dilatations, and special conformal transformations.
These transformations can be extended to $OSp(1,2M)$ acting on the
supermultiplet formed by scalar $b(X)$ and svector $f_\ga (X)$ and to
extended conformal supersymmetries $OSp(L,2M)$ acting on appropriate sets of
scalars and svectors \cite{BHS} (see  also section
\ref{Extended Supersymmetry}).

Note that we refer to the $Sp(2M)$
as to generalized conformal symmetry  not only because of
similarity with the usual conformal symmetry but also
because, as shown in section \ref{Towards any M}, it  extends
the usual conformal symmetry $SO(d,2)$ acting in the theory.
(For example, for the case of $d=4$ $(M=4)$ this is in accordance
with the well-known fact that $o(4,2)\sim su(2,2) \subset sp(8)$.)
An important feature of the equations (\ref{oscal}) and (\ref{ofer})
is that they do not contain any metric tensor. As a result, the
interpretation of $Sp(2M)$ as a conformal symmetry associated with some
metric  re-scaling may not necessarily be relevant for the full theory
in $\M_M$. In particular, the inversion $R$ (\ref{inv}) is defined
without any metric tensor, as opposed to the usual inversion
$x^i \to \f{x^i}{x^j x^k \eta_{jk}}$ where $\eta_{jk}$ is the
Minkowski metric tensor.

The rest of the paper is organized as follows. In  section
\ref{Propagation} we solve the equations (\ref{oscal}) and (\ref{ofer})
by means of  Fourier transform and analyze some particular solutions
associated with Green functions. The causal structure of
the generalized space-time $\M_M$ is investigated in section
\ref{Casual Structure} where the concepts of global Cauchy surface
and time are defined. The generalized Lorentz transformations
are discussed in section \ref{Lorentz}.
The problem of localizability of fields in $\M_M$
is studied in section
\ref{Field Localizability} where the concept of local Cauchy
bundle with the usual space as base manifold $\gs$ is introduced. The
particular cases of $M=2$ and $M=4$ are considered in detail in
subsections \ref{M=2} and \ref{M=4}. It is also shown in \ref{M=4}
that the $4d$ electro-magnetic duality transformations along with their
extension to higher spins identify with certain generalized Lorentz
transformations.
Quantization of the free fields in the generalized space-time is
performed in section \ref{Quantization} where the positive-definite
Hilbert space of one-particle states is built, $\D$ functions and Green
functions are found and the microcausality is analyzed including the
analysis of the spin-statistics relationship. A structure of the
generalized space-time for higher $M$ is discussed in section
\ref{Towards any M} with the emphasize on the key role of Clifford
algebras in the definition of the space $\gs$
for generic $M$. Generalized electro-magnetic duality is identified with
the subgroup of the generalized Lorentz symmetry that leaves invariant
a local Cauchy surface  $\gs$. Particular attention is paid to the
$M=8$ case that corresponds to a $6d$ chiral higher spin
theory. The cases of $M=16$ and $M=32$ corresponding to some
$10d$ and $11d$ theories are also discussed in section
\ref{Towards any M}. A $osp(2L,2M)$ invariant superspace extension of the
proposed equations is formulated in section \ref{Extended Supersymmetry}.
Coset constructions for the compactified generalized
(super)spaces are given in section \ref{CM}.
A summary of the obtained results and
outlook is the content of section \ref{Conclusions}.

\section{Classical Solutions}
\label{Propagation}

The equation (\ref{oscal}) admits a solution of the form
\be
\label{pulse}
b(X) = \phi (\xi_\ga\xi_\gb   X^{\ga\gb})\,
\ee
with an arbitrary function of one variable $\phi (z)$ and
constant parameters  $\xi_\ga$. Such solutions are
analogous to the plane wave
light-like solutions in the usual Minkowski space-time with coordinates
$x^n$ $(n=0\ldots  d-1 )$
\be
b(x) = \phi (x^n k_n )\,,\qquad k_n k^n =0\,. \nn
\ee
For $M=2$ and $d=3 $ the two formulas are  equivalent because $\xi_\ga$ defines
a light-like direction. For $d=4$, light-like wave vectors admit the
analogous twistor representation
$k_{a\dot{b}}=\xi_a\bar{\xi}_{\dot{b}}$ ($a,b\ldots=1,2;
\dot{a},\dot{b}\ldots =1,2)$.

To show that the set of solutions (\ref{pulse}) is complete
consider  Fourier transform. For a particular harmonic
\be
\label{expb}
b(X) = b_0 \exp i k_{\ga\gb}X^{\ga\gb}\,
\ee
 (\ref{oscal}) requires
\be
\label{kk}
k_{\ga\gb}k_{\gga\gd}=k_{\ga\gga}k_{\gb\gd}\,.
\ee
This is solved by the twistor ansatz
\be
\label{ktw}
k_{\ga\gb} =k \xi_\ga \xi_\gb
\ee
with an arbitrary commuting real svector $\xi_\ga$ and a factor $k$.
The equivalent statement is
that any non-zero matrix $k_{\ga\gb}$ satisfying (\ref{kk}) has rank 1.
For the proof it is enough to diagonalize the
symmetric matrix $k_{\ga\gb}$ by a $sl_M$ transformation to see that
the product of any two different eigenvalues is zero by  (\ref{kk})
at $\ga \neq \gb$. Modulo rescalings of $k$ and $\xi_\ga$,
there are two essentially different options in (\ref{ktw}) with
$k=1$ or $k=-1$. These correspond to the positive and negative
frequency solutions, respectively.

The situation with  the Dirac-like
equation (\ref{ofer}) is analogous because
it follows that $f_\ga (X)$ satisfies (\ref{oscal})
\bee
\Big (
\f{\p^2}{\p X^{\ga\gb} \p X^{\gga\gd}} &-&
\f{\p^2}{\p X^{\ga\gga} \p X^{\gb\gd}}\Big ) f_\gs (X)
\nn\\&=& \f{\p}{\p X^{\gs\gd}}\Big (
\f{\p}{\p X^{\ga\gb}            } f_\gga (X) -
\f{\p}{\p X^{\ga\gga}}       f_\gb (X)\Big ) =0\,.
\eee
A plane wave solution for the svector field $f_\ga (X)$ has a form
analogous to (\ref{pulse})
\be
\label{fpulse}
f_\ga (X) =\xi_\ga \phi (\xi_\gga\xi_\gb   X^{\gga\gb})\,.
\ee
The Dirac-like
equation (\ref{ofer}) requires  $f_{\ga}$
to be proportional to $\xi_\ga$. As a result, the harmonic svector
plane wave solution has the form
\be
\label{expf}
f_\ga (X) =f_{0}\xi_\ga \exp i k \xi_{\gga}\xi_{\gb} X^{\gga\gb}\,.
\ee
Scalar and svector therefore
have equal numbers of on-mass-shell degrees of freedom.

An important particular solution of the equation (\ref{oscal})
has the form
\be
\label{detsol}
b(X) = det^{-\half} |X-X_0 |\,,
\ee
where $X_0$ is any fixed point of $\M_M$. Setting $X_0 =0$,
one gets
\be
\f{\p}{\p X^{\ga\gb}} det^{-\half} |X|  =-\half
X_{\ga\gb} det^{-\half} |X|\,. \nn
\ee
Taking into account
\be
\f{\p}{\p X^{\ga\gb}} X_{\gga\gd} = -\half \Big ( X_{\ga\gd}X_{\gb\gga}  +
X_{\ga\gga}X_{\gb\gd} \Big )\nn
\ee
one obtains
\be
\f{\p^2}{\p X^{\ga\gb}\p X^{\gga\gd} } det^{-\half} |X|  =\f{1}{4} \Big
(X_{\ga\gb}X_{\gga\gd} + X_{\ga\gga}X_{\gb\gd}+ X_{\ga\gd}X_{\gb\gga} \Big )
det^{-\half} |X|\,. \nn
\ee
This expression is totally symmetric with respect to
$\ga,\gb,\gga,\gd$. As a result,
the antisymmetric part of the second derivative
of  $det^{-\half} |X|$ corresponding to the left hand side of
(\ref{oscal}) vanishes.

Analogously, the svector equation (\ref{ofer})
admits a solution
\be
\label{Xdet}
f_\ga (X) = (X-X_0)_{\ga\gb}\eta^\gb det^{-\half} |X-X_0 | \nn
\ee
with an arbitrary constant polarization svector $\eta^\gb$.

Formulas (\ref{detsol}) and (\ref{Xdet}) solve (\ref{oscal})
and (\ref{ofer}) at least in the regions where the matrix
$X^{\ga\gb} -X_0^{\ga\gb}$ is nondegenerate.
In section \ref{Quantization} we show that these solutions are related
to the Green functions of the scalar and svector in the generalized
space-time  and give more precise definition of their singularities.

\section{Causal Structure and Time}
\label{Casual Structure}

The equations (\ref{oscal}) and (\ref{ofer}) imply
propagation along  the generalized light-like directions
\be
\label{3}
\Delta X^{\ga\gb} = \eta^\ga \eta^\gb\,,
\ee
where $\eta^\ga$ is a twistor dual to $\xi_\gb$. The
sign choice on the right hand side of (\ref{3})
fixes a choice of the time arrow.

One way to reach this conclusion is to analyze the
characteristic equation for the front of discontinuity of a
field amplitude. Let $n_{\ga\gb}$ be proportional to
 the infinite part of the derivative
$\f{\p}{\p X^{\ga\gb}} b(X)$. For the front discontinuity of
co-dimension
one to  be compatible with the field equations the normal vector
$n_{\ga\gb}$ has to satisfy the equation analogous to (\ref{kk})
and therefore has to be of the form
\be
\label{nxi}
n_{\ga\gb} (z) = \pm \xi_\ga (z) \xi_\gb (z)\,,
\ee
where the coordinates $z$  parametrize the wave front.
For example, this formula is true for the solution
(\ref{pulse}) with the step-function $\phi(z)$.
The discontinuity fronts of the solutions
(\ref{detsol}) and (\ref{Xdet}) are
described by the surfaces of degenerate matrices
\be
det|X-X_0 | =0\,.          \nn
\ee
For a front of co-dimension one its normal vector is described by a
rank 1 matrix $n_{\ga\gb}$ satisfying
\be
(X^{\ga\gb} - X_0^{\ga\gb} )n_{\gb\gga} =0\,.  \nn
\ee
The svector $\xi_\ga$ in the formula (\ref{nxi}) identifies with
the null-vector of the matrix $X^{\ga\gb} - X_0^{\ga\gb} $.

Suppose that a light-like signal emitted from some point $X_0^{\ga\gb}$
of the generalized space-time reaches
some other point $X_1^{\ga\gb}=X_0^{\ga\gb} + \eta^\ga \eta^\gb$
switching on a new process that emits a signal in a different
light-like direction \footnote{Let us note that the assumption
that a process can be switched on locally
may or may not be true for particular dynamical equations.
In fact, as discussed in more detail in section
\ref{Field Localizability}, the equations
(\ref{oscal}), (\ref{ofer}) do not admit true localization in $\M_M$.
We do not expect this to
affect our  conclusions on the causal structure of the
generalized space-time, however.}.
Provided  this happens several times, any point
\be
\label{del}
\Delta X^{\ga\gb} =\sum_{i=0}^{M} \eta_i^\ga \eta_i^\gb\,,
\ee
can be reached
where $\eta_i^\ga$ is a complete set of contravariant svectors
dual to the complete set of covariant svectors $\xi_\ga^i$
\be
\xi^i_\ga \eta_i^\gb = \delta_\ga^\gb \q
\xi^i_\ga \eta_j^\ga = \delta^i_j \,.\qquad i,j = 1 \ldots M\,. \nn
\ee
Formula (\ref{del}) describes a general
positive semi-definite symmetric matrix $\Delta X^{\ga\gb}$.
Let us note that
analogous representation of positive semi-definite matrices in the
context of analysis of BPS states was used recently in \cite{BA}.
The authors of \cite{BA} called the elementary twistors $\eta_i^\ga$
preons.

We  see that the relativistic geometry that follows
from the equation (\ref{oscal}) identifies the future
cone $\C_{X_0}^+$ of a point $X_0$
with the set of matrices $X^{\ga\gb}$ such that
$\Delta X^{\ga\gb}=X^{\ga\gb}-X_0^{\ga\gb} $ is
positive semi-definite. Time-like vectors
are described by positive definite matrices
\be
\Delta X^{\ga\gb}\xi_\ga \xi_\gb >0\q \forall \xi_\ga\,. \nn
\ee
Light-like vectors identify with degenerate positive semi-definite
matrices
\be
det|\Delta X| =0\,,\qquad
\Delta X^{\ga\gb}\xi_\ga \xi_\gb \geq 0\q \forall \xi_\ga\,. \nn
\ee
We will distinguish between rank - $k$ light-like directions
described by matrices of rank $k$. The
concepts of time-like and rank - $k$ light-like vectors are invariant
under the generalized Lorentz group $SL_M$.

The equation (\ref{oscal}) describes propagation of signals along
the most degenerate light-like directions of rank 1.
Using the technique developed in \cite{BHS} one can work
out a form of the equations that describe propagation along
less degenerate light-like directions.  We will come back to
this question elsewhere.

The past cone $\C_{X_0}^-$ is defined analogously as the set of
negative semi-definite matrices
\be
(X^{\ga\gb}-X_0^{\ga\gb} ) \xi_\ga \xi_\gb \leq 0\q \forall \xi_\ga\,.
\nn
\ee
If $Y\in \C_{X}^+$ then $X \in \C_{Y}^-$ and
$2X-Y \in \C_{X}^-$.
Note that $\C_{X}^+$ is the convex cone: $\forall X_1 , X_2 \in
\C_{X}^+ , \lambda,\mu \in R^+\,,
\lambda X_1 +\mu X_2 \in \C_{X}^+$.

To define the concept of time let us first introduce a concept of
space-like global Cauchy surface as such a submanifold $\Sigma$
of some (generalized) space-time manifold $\M$ that

\noindent
$(i)$ $\forall X_1 , X_2 \in \Sigma $,
$X_1 \notin \C^\pm_{X_2}$ and $X_2 \notin \C^\pm_{X_1}$ for $X_1\neq X_2$.

\noindent
$(ii)$ any point of $\M$
belongs to either future or past cone of some point
of $\Sigma$. No point $Y\in \M$ can belong to the future
cone of some point $X_1 \in \Sigma$ and past cone of some
other point $X_2 \in \Sigma$.

The meaning of the definition of global Cauchy surface
is obvious: no pair of observers on $\Sigma$
are allowed to exchange causal signals, i.e. a global Cauchy surface
is space-like; emitting causal signals from a Cauchy surface, one can
reach any point in the future, and the
space-like global Cauchy surface can be reached by a signal from any
space-time point in the past\footnote{Note that this definition can be adjusted
to a particular type of signals by replacing in the condition $(ii)$
the future and past cones by their boundary of a particular type
(say, rank 1 for the case under consideration).  Such a
specification does not make difference at least for the particular
dynamical system we study.}. Note that for the particular case under
consideration with
$\M = \M_M$ being a linear space $R^{\half M(M+1)}$ and
convex future (past) cones
the second part of the requirement $(ii)$ is a consequence of $(i)$.

Provided that $\M$ admits a fibration into a set of
space-like global Cauchy surfaces
$\Sigma_t$ parametrized by some parameter(s) $t$, this defines the
concept of time(s).

Let $T^{\ga\gb}$ be some positive definite matrix.
The axioms $(i)$ and $(ii)$ are satisfied with the
space-like global Cauchy surfaces $\Sigma_t$ parametrized as
\be
\label{res}
X^{\ga\gb} \in \Sigma_t :\qquad X^{\ga\gb} = x^{\ga\gb}+tT^{\ga\gb}\,,
\ee
where the space coordinates $x^{\ga\gb}$ are
arbitrary $T-$ traceless matrices
\be
\label{xtr}
x^{\ga\gb}T_{\ga\gb}=0\,,\qquad T_{\ga\gb}T^{\gb\gga} =\delta_\ga^\gga\,.
\ee
Indeed, the difference of any two matrices
of the form (\ref{res}) at fixed $t$ is traceless and
therefore it is neither positive definite nor negative definite.
As a result, any two points of $\Sigma_t$  at some fixed $t$
are separated by a space-like interval.
The rest of the axioms is a consequence of the trivial decomposition
(\ref{res}) of a matrix into the sum of its trace and traceless parts.

An important output of this analysis is that the
generalized space-time $\M_M$ has just one evolution parameter
\be
\label{t}
t=\f{1}{M} X^{\ga\gb} T_{\ga\gb}\,.
\ee
The ambiguity in the choice of a positive definite
matrix $T^{\ga\gb}$ parametrizes the ambiguity in the choice of a
particular coordinate frame like in Einstein special
relativity: any two positive definite
matrices $T_{1,2}^{\ga\gb}$ with equal
determinants are related by some
generalized $SL_M$ Lorentz transformation. The dilatation
allows one to fix a scale of time in an arbitrary way.

Note that rank - $k$ light-cone time parameters can be defined analogously
with positive semi-definite matrices $T^{\ga\gb}$ in (\ref{res}).

\section{Generalized Lorentz Transformations}
\label{Lorentz}

Having defined the concept of time, we are now in a position
to analyze the  generalized Lorentz transformations
\be
X^{\prime\ga\gb} = a^\ga{}_\gga a^\gb{}_\gd X^{\gga\gd}\,.\nn
\ee
Here $a^\ga{}_\gga$ is an $SL_M$ matrix
\be
\label{deta}
det\Big |a^\ga{}_\gga \Big |  =1\,.
\ee

The space symmetry subalgebra of the Lorentz-like group
$SL_{M}$ consists of the elements that
leave invariant the positive definite
symmetric matrix  $T^{\ga\gb}$ associated with time
\be
\label{rotin}
T^{\ga\gb} = a_o{}^\ga{}_\gga a_o{}^\gb{}_\gd T^{\gga\gd}\,.
\ee
It is isomorphic to the compact group $SO(M)$ being the maximal
compact subgroup of $SL_M$. $SO(M)$ is the analog of the usual space
rotations  $SO(d-1)$ of the $SO(d-1,1)$ Lorentz invariant space-time.
The dimension of the coset space
$SL_M /SO(M)$ equals to the number of the space
(traceless) matrix coordinates
$\half M(M+1) -1$. As a result, very much as for the usual
Lorentz transformations, the parameters of the generalized Lorentz
group turn out to be associated with the space symmetry and the
generalized velocities.

Let us introduce the following quantities
\be
u^{\ga\gb} (a) = a^\ga{}_\gga a^\gb {}_\gd T^{\gga\gd} - T^{\ga\gb} \gga (a)\,,
\nn
\ee
\be
r_{\gga\gd} (a) = a^\ga{}_\gga a^\gb {}_\gd T_{\ga\gb} - T_{\gga\gd} \gga (a)\,,
\nn
\ee
\be
\gga (a) =\f{1}{M} a^\ga{}_\gga a^\gb {}_\gd T^{\gga\gd}T_{\ga\gb} \,\nn
\ee
defined in such a way that
\be
u^{\ga\gb} T_{\ga\gb} =0\,,\qquad
r_{\ga\gb} T^{\ga\gb} =0\,. \nn
\ee
According to these definitions, $u^{\ga\gb} (a)$ and $r_{\ga\gb} (a)$
parametrize the right and left
coset spaces $SL_M /SO(M)$, respectively,
\be
u^{\ga\gb} (a a_o)=u^{\ga\gb} (a)\,,\qquad
r_{\ga\gb} ( a_o a)=r_{\ga\gb} (a)\,,\nn
\ee
where $a_o \in SO(M)$ is any generalized
 space rotation satisfying (\ref{rotin}).
The parameter $\gga (a)$ is left and right invariant
\be
\gga (a a_o)=\gga (a_o a) = \gga (a)\,. \nn
\ee
It is not independent, but can be uniquely expressed in terms
of either $u^{\ga\gb}$ or $r_{\ga\gb}$.  To see this, consider the
positive definite symmetric matrices
\be
U^{\ga\gb} (a) = a^\ga{}_\gga a^\gb {}_\gd T^{\gga\gd} \,,\qquad
R_{\gga\gd} (a) = a^\ga{}_\gga a^\gb {}_\gd T_{\ga\gb} \, \nn
\ee
and their characteristic equations
\be
\label{VR}
det \Big | U^{\ga\gb} (a) -\lambda T^{\ga\gb} \Big | =0\,,\qquad
det \Big | R_{\ga\gb} (a) -\lambda T_{\ga\gb} \Big | =0\,.
\ee
It is elementary to see that, the equations on $U$ and $R$
in (\ref{VR}) are equivalent and, therefore, have the same
sets of eigenvalues $\lambda_i$.
Since $U^{\ga\gb}, R_{\ga\gb}$ and $T_{\ga\gb}$ are positive definite,
all eigenvalues are strictly positive
\footnote{Let us note that the construction of the representatives of the
left and right coset spaces $U^{\ga\gb}$ and $R_{\ga\gb}$ is
analogous to the construction of the metric tensor in the frame
formulation of gravity ($a^\ga{}_\gb$ and $T^{\ga\gb}$ are analogues
of the frame field and flat metric, respectively). The eigenvalues
(\ref{vi}) parametrize the double coset space $SO(M)\setminus SL_M/SO(M)$.}
\be
\lambda_i > 0\,. \nn
\ee
{}From  (\ref{deta}) it follows that
\be
\label{prodl}
\prod_{i=1}^M \lambda_i = 1\,.
\ee
The eigenvalues of $u^{\ga\gb}$ and $r_{\ga\gb}$ are
\be
\label{vi}
u_i = \lambda_i - \f{1}{M}\sum_{j=1}^M \lambda_j\,.
\ee
Taking into account that
\be
\label{gbl}
\gga (a) = \f{1}{M}\sum_{j=1}^M \lambda_j\,
\ee
(\ref{prodl}) acquires the form
\be
\label{prodv}
\prod_{i=1}^M (u_i+\gga) = 1\,.
\ee
This equation allows one to express $\gga$ in terms of $u_i$ uniquely
at the condition that all factors $(u_i+\gga)$ are strictly positive
(the function $\prod_{i=1}^M (u_i+\gga)$ is monotonic in $\gga$
in the region where the factors $(u_i+\gga)$ are positive). This proves
that $\gamma$ expresses in terms of both $r_{\ga\gb}$ and $u^{\ga\gb}$.
As a consequence of the well-known inequality between the
arithmetic and geometric averages,
from (\ref{prodl}) and (\ref{gbl}) it follows that
\be
\gga (u) \geq 1\,. \nn
\ee

Now we rewrite the generalized Lorentz transformations
in terms of the decomposition (\ref{res})
\be
\label{xpr}
x^{\prime\ga\gb} = a^\ga{}_\gga a^\gb{}_\gd x^{\gga\gd}
-\f{1}{M}T^{\ga\gb}  r_{\gga\gd} x^{\gga\gd} +t u^{\ga\gb}\,.
\ee
\be
\label{tpr}
t^\prime = \f{1}{M}r_{\ga\gb} x^{\ga\gb} +\gga t\,.
\ee
In this analysis we assume that the space-time
decomposition (\ref{res}) is defined with respect to the same matrix
$T^{\ga\gb}$ in the both frames. If one would rotate the matrix
$T^{\ga\gb}$ by the same Lorentz transformation the description
in the two systems would be identical. For example the $3d$ coordinates
in the decomposition $X^{\ga\gb} = t I^{\ga\gb} +x^1 \gs_1^{\ga\gb} +
x^2 \gs_3^{\ga\gb}$ are defined for the fixed basis matrices
$T^{\ga\gb}=I^{\ga\gb}$ and  $\gs_i^{\ga\gb}$.

{}From (\ref{tpr}) one observes that $\gga (u)$ is
a generalization of the relativistic Lorentz factor that relates
the time scales in the two systems.
The velocity of the origin of coordinates $x^{\ga\gb}$ of
the unprimed generalized inertial coordinate frame
with respect to the primed one is
\be
v^{\ga\gb} =       \f{ u^{\ga\gb}}{ \gga (u)}\,. \nn
\ee
The eigenvalues
$v_i$ of the traceless matrix $v^{\ga\gb}$  are
\be
v_i = \f{M\lambda_i}{\sum_{j=1}^M \lambda_j}-1 \,. \nn
\ee
{}From this formula it follows that the eigenvalues
satisfy the restrictions
\be
\label{gc}
M-1 \geq v_i\geq -1\,.
\ee
If  one of the eigenvalues $\lambda_i \to \infty$,
$v_i$ saturates the upper bound while $v_j$ at $j\neq i$ saturate
the lower bound. For these limiting cases the relativistic factor
$\gga (v)\to \infty$. It is elementary to see using (\ref{deta}) that
this is a general phenomenon: if some $v_i \to M-1$ then $v_j \to -1$
$j\neq i$ and $\gga (v) \to \infty$. Moreover, $\gga (v) \to \infty$
whenever at least one of the eigenvalues $v_j \to -1$ (while the
upper bound may not be saturated).

The condition (\ref{gc})  is a generalization of the Lorentz
geometry restriction that one system cannot move with respect
to another with the speed exceeding the speed of light.
Let us note that there is no symmetry
$v_i \to -v_i$ in the generalized geometry because eigenvalues
$v_i$ are invariants of the space symmetry group $SO(M)$. In the
usual Lorentzian geometry a sign of the velocity vector can be
changed by a space rotation. A generalization of this symmetry
to  the generalized geometry is the symmetry  under
permutations of eigenvalues $v_i \leftrightarrow v_j$. Note that,
according to the realization of Lorentz boosts (\ref{boo})
 of section \ref{Towards any M}, usual Lorentz
transformations identify with such  generalized
Lorentz transformations that there are only two different eigenvalues
among $v_i$ which, therefore, can differ only by sign, thus belonging
to the interval $1\geq v_i \geq -1$. In particular, this is obviously
the case for $M=2$.

The transform from the primed frame to the unprimed one is
described by the inverse $SL_M$ transform and, therefore is
characterized by the replacement $\lambda_i \to \lambda_i^{-1}$. Note that
the relativistic factors $\gga (v)$ of the direct and inverse transforms
are not necessarily equal to each other. However,
because of (\ref{deta}) they tend to infinity simultaneously, i.e.
if a system $A$ is ultrarelativistic with respect to the system $B$,
then the system $B$ is ultrarelativistic with respect to $A$.
A system at rest is characterized by $v_i =0$.

\section{Field Localizability}
\label{Field Localizability}

{}The definition of a global Cauchy surface suggests that
it is enough to know values of the fields
along with some their time derivatives on a global Cauchy surface to
fix a particular solution in the whole
generalized space-time $\M_M$. This is certainly true. The question is
what is a set of initial data that can be fixed arbitrarily
to determine the time evolution
of fields. Because a single field $b(X)$ satisfies
the system of equations (\ref{oscal}), some of
these equations play a role of constraints on the global Cauchy surface thus
restricting a possible choice of the initial data. This is
analogous to the usual constraint dynamics. For example the Gauss
low constraint $\p_i E^i =0$ in the pure electrodynamics restricts
initial data for the electric field $E^i$.

Indeed, using the decomposition (\ref{res}) we have
\be
\f{\p}{\p X^{\ga\gb}}=\f{1}{M}T_{\ga\gb}\f{\p}{\p t}+
\f{\p}{\p x^{\ga\gb}}\,,\nn
\ee
where the space coordinates $x^{\ga\gb}$ are traceless in the sense
of (\ref{xtr}).
{}From (\ref{oscal}) one derives the wave equation
\be
\Big (\f{\p^2}{\p t^2} -\f{M}{M-1} T^{\ga\gga}T^{\gb\gd}
\f{\p^2}{\p x^{\ga\gb}\p x^{\gga\gd}} \Big ) b(X) =0 \nn
\ee
and constraints
\bee
\label{cons}
&{}&\Big (T_{\ga\gb} \f{\p}{\p x^{\gga\gd}}+
T_{\gga\gd} \f{\p}{\p x^{\ga\gb}}\Big ) p(X) \\
&+&\Big ( M\f{\p^2}{\p x^{\ga\gb}\p x^{\gga\gd}}
+\f{1}{M-1}T_{\ga\gb} T_{\gga\gd} T^{\nu\mu} T^{\rho\sigma}
\f{\p^2}{\p x^{\nu\rho}\p x^{\mu\sigma}}
\Big ) b(X)\nn -(\gb\leftrightarrow\gga )=0\,,
\eee
where
$
p(X) = \f{\p}{\p t} b(X)\,.
$
Note that the totally $T-$transversal part of the constraints
(\ref{cons}) is independent of the momentum $p$ thus imposing
constraints on the space derivatives of $b(X)$.

The constraints restrict initial data for the field $b(x,t_0)$
and its first derivative $p(x, t_0)$ on the global Cauchy surface. In
particular, it is not possible to choose
initial data localized at some point $x_0$
with $b(x, t_0)\sim \delta(x-x_0 )$,
$p(x, t_0)\sim \delta(x-x_0 )$. This is why we say that the fields
$b(X)$ and $f_\ga (X)$ are not localizable in the generalized space-time
$\M_M$.

This conclusion is not surprising because, as shown
in  section \ref{Propagation},
the generic solutions of the equations (\ref{oscal}) and (\ref{ofer})
have the form
\be
\label{bfo}
b (X) =\f{1}{\pi^{\f{M}{2}}}
\int d^M\xi\, \Big ( b^+ (\xi ) \exp i  \xi_{\ga}\xi_{\gb} X^{\ga\gb}
+b^- (\xi ) \exp -i  \xi_{\ga}\xi_{\gb}X^{\ga\gb} \Big ) \,,
\ee
\be
\label{ffo}
f_\gga (X) =\f{1}{\pi^{\f{M}{2}}}
\int d^M\xi\, \xi_\gga
\Big ( f^+ (\xi ) \exp i  \xi_{\ga}\xi_{\gb} X^{\ga\gb}
+ f^- (\xi ) \exp -i  \xi_{\ga}\xi_{\gb}X^{\ga\gb} \Big ) \,.
\ee
Both for the scalar $b(X)$ and svector $f_\ga (X)$,
the space of solutions is parametrized by two
functions of $M$ variables $\xi_\ga$. Because odd functions
$b^\pm (\xi)$ and even functions
$f^\pm (\xi)$ do not contribute to (\ref{bfo})
and (\ref{ffo}), respectively, we require \be
\label{par} b^\pm (\xi)=b^\pm (-\xi) \,,\qquad
f^\pm (\xi)=-f^\pm (-\xi)\,.  \ee
The integration in (\ref{bfo}) and (\ref{ffo}) is thus carried out over
$R^M / Z_2$. The origin of coordinates $\xi_\ga =0$ is invariant under the $Z_2$
reflection $\xi_\ga\to -\xi_\ga$ and therefore is a singular
point of the conical orbifold $R^M / Z_2$.

Using the ambiguity in $b^\pm (\xi) $ and $f^\pm (\xi )$ it may be
possible to achieve localization in at most $M$ coordinates that,
generically, is much less than the dimension of the global Cauchy surface
$dim(\Sigma ) = \half M(M+1) -1$.

Let us now introduce the concept of local Cauchy bundle. Naively,
one might try to identify it with some submanifold $\gs$ of the global
Cauchy surface $\Sigma$ such that, fixing a certain number of
arbitrary functions on $\gs$, the constraints (\ref{cons})
reconstruct the initial data on $\Sigma$. This would imply that
the fields would allow a true localization on $\gs$ rather than on
$\Sigma$. Since local observers can only distinguish between local
events such a picture would mean that $\gs$ is a visualization
of the global Cauchy surface by means of a particular field
dynamics under consideration.

This idea is basically true with the correction
that a number of space
coordinates $d-1$ that allow true localization
may be even less than $M$.
A relevant object called local Cauchy bundle $E$
is a $M-$dimensional fiber bundle
over an appropriate $d-1-$dimensional base manifold
$\sigma \in \Sigma $ called local Cauchy surface and
treated as the space manifold. The local Cauchy surface
$\gs$ is a submanifold of $\M_M$. The space-time manifold is
$R \times \sigma \subset \M_M$ where $R$ is the time axis.
Note that the local Cauchy bundle $E$ is not necessarily
a submanifold of $\M_M$. Since the dynamics
in the generalized space-time was argued in section
\ref{Casual Structure} to be compatible with the causal structure
of the generalized space-time, the projection of this dynamics to the
space-time $R\times \sigma$ that admits localization of
the fields is expected to be compatible with the microcausality
principle, i.e. the restriction of the Green functions to some
local Cauchy surface $\gs$ is expected to vanish for the space-like
separated regions. We will come back to this point in section
\ref{Quantization}.

Let us stress that different types of fields that may live in
the same generalized space-time may require local Cauchy bundles
of different dimensions thus
providing different visualizations of the same generalized space-time.
This phenomenon is analogous to the fact
known after Dirac \cite{Dirac} that the singleton field lives on the
boundary of $AdS_4$ while all other $AdS_4$ fields live in the bulk.
The parallels with the ideas of holography \cite{Hol,AdS/CFT} and
brane dynamics are self-suggestive either.
Note that even for the same dynamical system
 a choice of a particular local Cauchy
bundle $E$ may {\it a priori} be not unique. Different choices of $E$
can lead to different descriptions of the same dynamical system.
Although being equivalent, the descriptions in terms of
different local Cauchy bundles may look differently and,
in fact, describe dual versions of the same model that has
a uniform description in the full generalized space-time.

Let us consider some examples.

\subsection{M=2}
\label{M=2}

For $M=2$, the generalized
space-time reduces to the usual $3d$ space-time geometry
while the equations (\ref{oscal}) and (\ref{ofer}) are equivalent to
the usual massless equations for $3d$ scalar and spinor.
There are no constraints (\ref{cons}) for $M=2$. Let us
show that, for this case, the representation (\ref{bfo}) allows for
the usual field localizability.

We set
\be
T^{\ga\gb} = \delta^{\ga\gb}\,, \nn
\ee
\be
X^{\ga\gb} = t \delta^{\ga\gb} +x^1 \sigma_1^{\ga\gb}
+x^2 \sigma_3^{\ga\gb} \,,  \nn
\ee
where $\sigma_{1,3}^{\ga\gb}$ are the two traceless symmetric Pauli
matrices having unit square. Restriction of the
solution (\ref{bfo}) to the global Cauchy surface $t=0$ gives
\be
\label{2bfo}
b (x,0) =\f{1}{\pi }
\int d^2\xi\, \Big (
b^+ (\xi )\exp  i (k_1 x^1 +k_2 x^2 )
+b^- (\xi )\exp - i (k_1 x^1 +k_2 x^2 )
\Big ) \,,
\ee
where
\be
\label{k2}
k_1 =\xi_1^2 - \xi_2^2 \,,\qquad k_2 = 2  \xi_1 \xi_2 \,.
\ee
The combinations of the integration variables (\ref{k2})
map $R^2 /Z_2$ on $R^2$. This map is bijective with the
expected singularity at $\xi_\ga =0$
\be
dk_1 \wedge dk_2 = 4\Big (\xi_1^2 + \xi_2^2 \Big )d\xi_1
\wedge d\xi_2\,. \nn
\ee
Note that
\be
\xi_1^2 + \xi_2^2  = \sqrt{k_1^2 +k_2^2}\,. \nn
\ee

Setting
\be
b^\pm (\xi ) =\f{1}{2 \pi} \Big (\xi_1^2 + \xi_2^2 \Big )
\exp{\mp ik_i x_0^i } \nn
\ee
one obtains
\be
b (x,0)
=\delta(x^i - x^i_0) \qquad \f{\p}{\p t}  b (x,t)\Big |_{t=0} =0 \,.
\nn
\ee
Analogously, setting
\be
b^\pm (\xi ) =\pm \f{1}{2\pi i} \exp{\mp ik_i x_0^i }\,,\nn
\ee
we obtain
\be
b (x,0) =0\,,\qquad
\f{\p}{\p t}  b (x,t)\Big |_{t=0} =\delta(x^i - x^i_0) \,. \nn
\ee

Thus, the twistor parametrization  (\ref{bfo})
of the solutions of the $3d$ field equations is equivalent to
the standard Minkowski parametrization with the integration
measure $d^3 k \delta (k^2)$. As expected,
the initial data for $b(X)$ and its first time derivative can be
fixed in an arbitrary way on the global Cauchy surface $\Sigma$.
The analysis of the fermionic solutions (\ref{ffo}) is analogous.

\subsection{M=4}
\label{M=4}

The case of $M=4$ was argued in \cite{BHS}
to be equivalent to the free field equations of $4d$ conformal
fields of all spins. The situation here is more interesting
because the generalized space-time is ten-dimensional while
the physical space-time is four-dimensional. Having four twistor
integration parameters in (\ref{bfo}), a submanifold
of the global Cauchy surface that admits localization of fields
can be at most four-dimensional.
Let us show that the local Cauchy bundle is $R^3 \times S^1$ where
$\gs = R^3$ is the usual Cauchy surface of the $4d$ Minkowski
space-time while the $S^1$ harmonics
distinguish between spins of $4d$ conformal fields.

Using the language of two-component complex spinors we
set
\be
\label{4dec}
X^{\ga\gb} = \Big ( x^{ab} , \X^{a\da} , \overline{x}^{\da\db} \Big )
\ee
with the convention that the complex conjugation transforms
dotted indices $a,b = 1,2$ to the undotted ones
$\da,\db=1,2 $ and vice versa so that $\overline{x}^{\da\db} $ is
complex conjugated to $x^{ab}$ while  $\X^{a\da}$ is hermitian.
We choose the global Cauchy fibration as follows
\be
\label{4dfib}
X^{a\da}=\X^{a\da} =t\delta^{a\db} +x^i \sigma_i^{a\db}\,,\qquad
X^{ab} =x^{ab}\,,\qquad \overline{X}^{\da\db} =\overline{x}^{\da\db} \,,
\ee
where $\sigma_i^{a\db}$  are hermitian traceless Pauli matrices
normalized to have unit square. Having four integration variables
$\xi_\ga$ we can try to localize some four coordinates from among
those nine that parametrize the global Cauchy surface.

The coordinates of the local Cauchy surface $\sigma$ can be identified with
the three space coordinates $x^i$ of the usual $4d$ space-time.
Let us use the complex notation for the space coordinates in the
$1:2$ plane
\be
x=x^1+ix^2\,,\qquad \bar{x} = x^1-ix^2 \,. \nn
\ee
The combinations of svectors $\xi_\ga$ dual to the coordinates $x^i$
\be
\label{k3}
k_3 =\xi_1\bar{\xi}_{\dot{1}}-\xi_2\bar{\xi}_{\dot{2}} \,,\qquad
\overline{k} =2 \xi_1 \bar{\xi}_{\dot{2}} \,,\qquad
k=2 \bar{\xi}_{\dot{1}} \xi_{2}
\ee
map $R_4 /Z_2$ on $R_3$, i.e. $k_3,k$ and $\overline{k}$ can take
arbitrary values. The leftover ambiguity in the integration variables
$\xi_\ga$ for fixed $k_i$ is the overall phase factor
$\xi_\ga \to \exp{\half i\varphi}\, \xi_\ga$, $ \varphi \in [0,2\pi ]$.
(Recall that $\xi_\ga$ is identified with $-\xi_\ga$.) We set
\be
\label{cyc}
\exp{i\phi} = 2 \f{\xi_1 \xi_2}{k}\,,\qquad
\exp{-i\phi} = 2 \f{\overline{\xi}_{\dot{1}}
\overline{\xi}_{\dot{2}}}{\overline{k}}\,.
\ee
For the integration  measure  we obtain
\be
d(\xi_1\bar{\xi}_{\dot{1}}-\xi_2\bar{\xi}_{\dot{2}})\wedge
d(\xi_1 \bar{\xi}_{\dot{2}}) \wedge d(\bar{\xi}_{\dot{1}} \xi_{2} )
\wedge d( \xi_1 \xi_{2} )
=2\xi_1 \xi_2
(\xi_1\bar{\xi}_{\dot{1}}+\xi_2\bar{\xi}_{\dot{2}})
d\xi_1\wedge d\bar{\xi}_{\dot{1}}\wedge d\xi_2\wedge d\bar{\xi}_{\dot{2}}\,.
\nn
\ee
This is equivalent to
\be
dk_3 \wedge dk \wedge d\bar{k} \wedge d \phi
=-8 i (\xi_1\bar{\xi}_{\dot{1}}+\xi_2\bar{\xi}_{\dot{2}})
d\xi_1\wedge d\bar{\xi}_{\dot{1}}\wedge d\xi_2\wedge d\bar{\xi}_{\dot{2}}\,.
\nn
\ee
The map (\ref{k3}), (\ref{cyc}) from $R^4 /Z_2$ associated
with the integration variables $\xi_\ga$ to $R^3\times S^1$ described
by the variables $k_i$, $\phi$ is non-degenerate except the expected
singularity at $\xi_\ga =0$. Note that
\be
\xi_1\bar{\xi}_{\dot{1}}+\xi_2\bar{\xi}_{\dot{2}} =
\sqrt{k \bar{k}+k_3^2} =
\sqrt{k_1^2 +k_2^2+k_3^2 }\,.\nn
\ee

The integration over three noncompact momentum variables is expected
to localize three space coordinates associated with the
local Cauchy surface $\gs = R^3$.
Using the ambiguity in the cyclic momentum variable
$\phi$ one can  distinguish between different
angular dependencies on  the complex coordinate
\be
z=x^{12}\,,\quad\bar{z} = \overline{x}^{\dot{1}\dot{2}}\,.\nn
\ee
Fixing $|z|=r$, this is equivalent to considering functions on
$S^1$.

Setting all other coordinates to zero, consider the following
restriction of the solution of the field equations
\bee
\label{4dca}
\ls b(x^3,x,\bar{x} ,z, \bar{z}) =&{}&\ls\ls
\f{1}{\pi^{2}}\ls\,\,
\int d^4\xi\, \Big (
            b^+ (\xi ) \exp i(x^3 k_3 +x\bar{k}  +\bar{x} k
+ z {k} \exp {i \phi}+\bar{z} \bar{k} \exp {-i \phi})\nn\\
&{}&\ls\ls\ls\ls +
b^- (\xi ) \exp -i(x^3 k_3 +x\bar{k}  +\bar{x} k
+ z {k} \exp {i \phi}+\bar{z} \bar{k} \exp {-i \phi})\Big )\,.
\eee
{}From this expression with the coefficients $b^\pm (\xi )$ of the form
\be
\label{4bf}
b^\pm (\xi ) =\f{1}{8\pi^2 }
(\xi_1\bar{\xi}_{\dot{1}}+\xi_2\bar{\xi}_{\dot{2}}) f(\exp - i\phi )
 \exp \mp i(x^3_0 k_3 +x_0\bar{k} +\bar{x}_0k )
\,,
\ee
or
\be
\label{4bft}
b^\pm (\xi ) =\pm\f{1}{8\pi^2 i } f(\exp - i\phi )
 \exp \mp i(x^3_0 k_3 +x_0\bar{k} +\bar{x}_0k )
\,,
\ee
where $f(w^{-1})$ is some Laurant polynomial, it is clear that,
for $r\neq 0$, the
solution (\ref{4dca}) is not localized at $x^i = x_0^i$
because it contains an infinite power series in the derivatives
of $\delta(x^i - x^i_0 )$ with higher powers of $r^2=z\bar{z}$
in front of the higher derivatives of the delta-functions. It is
therefore impossible to achieve further localization on
$R^3\times S^1\subset\M_M$ using
the ambiguity in the cyclic momentum variable $\phi$.  This problem can be
avoided by taking the limit $r\to 0$ and keeping the leading terms of a
given phase. This is equivalent to neglecting all terms that contain
$z\bar{z}$, i.e., to considering analytic or antianalytic functions in $z$.
The picture with $r=0$ is most appropriate physically because the whole
dynamical information is then localized at some point of the space $R^3$
equipped with an auxiliary $S^1$ that does not affect the analysis of
locality and causality. This is why the $M=4$ local Cauchy bundle
$E=R^3 \times S^1$ is not a submanifold of the global Cauchy surface.

Indeed, from (\ref{4bf}) one obtains that
\be
\label{bl}
b(x^3,x,\bar{x},z,0) \Big |_{t=0} =\f{1}{2\pi i}
\oint \f{dw}{w} f(w^{-1}) \delta (x^3 -x^3_0 ) \delta (x -x_0  ) \delta
(\bar{x}-\bar{x}_0 + zw )\,,
\ee
\be
\label{bla}
b(x^3,x,\bar{x},0,\bar{z}) \Big |_{t=0} = \f{1}{2\pi i}
\oint \f{dw}{w}f(w^{-1}) \delta (x^3 -x^3_0 )
\delta (x -x_0+\bar{z}w^{-1} ) \delta
(\bar{x}-\bar{x}_0  )\,,
\ee
\be
\f{\p}{\p t} b(x^3,x,\bar{x},z,\bar{z}) \Big |_{t=0}=0\,. \nn
\ee
Analogously, from (\ref{4bft}) one obtains that
\be
b(x^3,x,\bar{x},z,\bar{z})\Big |_{t=0} =0\,, \nn
\ee
\be
\label{tbl}
\f{\p}{\p t}b(x^3,x,\bar{x},z,0) \Big |_{t=0} = \f{1}{2\pi i}
\oint \f{dw}{w} f(w^{-1}) \delta (x^3 -x^3_0 ) \delta (x -x_0  ) \delta
(\bar{x}-\bar{x}_0 + zw )\,,
\ee
\be
\label{tbla}
\f{\p}{\p t} b(x^3,x,\bar{x},0,\bar{z}) \Big |_{t=0} =  \f{1}{2\pi i}
\oint \f{dw}{w}f(w^{-1}) \delta (x^3 -x^3_0 )
\delta (x -x_0+\bar{z}w^{-1} ) \delta
(\bar{x}-\bar{x}_0  )\,.
\ee

A power of a polynomial in $z$ or $\bar{z}$ equals to the spin
of the $4d$  field associated with a particular $S^1$ harmonic.
Therefore, the higher spin is
the more derivatives of the space delta-function
appear  in the equation (\ref{bl}).  This property manifests the fact that
the conformal higher spin fields contained in the generating function
$b(X)$ admit interpretation as order-$s$ derivatives of the
dynamical potential fields like Maxwell field strength for spin 1,
Weyl tensor for spin 2 etc (for more details see
\cite{Ann,Gol}). The fundamental higher spin
gauge fields (potentials) are expected to allow
$\delta - $functional localization without extra derivatives
on the $3d$ local Cauchy surface for any spin.
The analysis of the fermionic equation (\ref{ofer}) is analogous.

The analysis of this subsection proves the conjecture of \cite{BHS}
that the system of equations (\ref{oscal})
and (\ref{ofer}) at $M=4$ is equivalent to the infinite
set of $4d$ equations of motion for massless fields of all spins.
This fact is not trivial because the consideration of \cite{BHS}
was essentially local in the extra six coordinates  of the generalized
space-time. To summarize, what happens is that the independent
degrees of freedom of the fields satisfying the equations
(\ref{oscal}) and (\ref{ofer}) in $\M_M$ live on a four-dimensional
local Cauchy bundle $E=R^3\times S^1$
with the base manifold $R^3$ identified with the
usual space and the fiber $S^1$ giving rise to the infinite tower of spins.
The dependence on the extra five coordinates of the global Cauchy surface
is reconstructed uniquely by the constraints (\ref{cons}). As a result,
propagation of the fields $b(X)$ and $f_\ga(X)$ in the generalized space-time
is equivalent to the propagation of local higher spin fields in the $4d$
space-time supplemented with one additional coordinate for spin.

Let us show how one can see this directly from  the equations
(\ref{oscal}) and (\ref{ofer}) focusing for definiteness on
the bosonic case. According to (\ref{4dec}) we write
$b = b(\X^{a\db} , x^{a b} , \bar{x}^{\da\db})$. Eq.(\ref{oscal})
decomposes into the three types of equations
\be
\label{XX}
\Big (\f{\p^2}{\p x^{ab} \p \bar{x}^{\da\db}}
- \f{\p^2}{\p \X^{a \da} \p {\X}^{b\db}} \Big ) b (\X, x ,\bar{x} )=0\,,
\ee
\be
\label{xx}
\Big (\f{\p^2}{\p x^{ab} \p {x}^{ c d}} - \f{\p^2}{\p x^{ac} \p {x}^{ b d}}
\Big ) b ( \X , x , \bar{x} ) =0\,,\quad\!\!
\Big (
\f{\p^2}{\p \bar{x}^{\da\db} \p \bar{x}^{ \dc \dd}} -
\f{\p^2}{\p \bar{x}^{\da\dc} \p \bar{x}^{ \db \dd}} \Big )
b ( \X , x ,\bar{x} ) =0\,,
\ee
\be
\label{Xx}
\!\!\!\!\Big (\f{\p^2}{\p x^{ab} \p {\X}^{c\da}}
- \f{\p^2}{\p x^{a c} \p {\X}^{b\da}} \Big ) b (\X, x, \bar{x} )=0,\quad\!\!
\Big (\f{\p^2}{\p \bar{x}^{\da\db} \p {\X}^{c\dd}}
- \f{\p^2}{\p \bar{x}^{\da \dd} \p {\X}^{c\db}} \Big ) b (\X, x, \bar{x}
)=0.
\ee

The equation (\ref{XX}) has two consequences. First, it determines
the dependence on the complex coordinates $x^{ab}$ and their conjugates
$\bar{x}^{\da\db}$ in terms of $\X$-derivatives of the analytic and
antianalytic functions $c(\X,x) =b(\X,x,0)$ and
$\bar{c}(\X,\bar{x}) = b(\X,0,\bar{x})$.
Second, its part antisymmetric in $a,b$ (equivalently, $\da,\db$)
implies the massless Klein-Gordon equation
\be
\epsilon^{ab} \epsilon^{\da\db}
\f{\p^2}{\p \X^{a \da} \p {\X}^{b\db}} b (\X, x , \bar{x} )=0\,, \nn
\ee
where $\epsilon^{ab}$ and $\epsilon^{\da\db}$ are the $2\times 2$ antisymmetric
symbols.

The equations (\ref{xx}) imply that the coefficients of the expansion of
$b (\X, x, \bar{x} )$ in powers of $x$ and $\bar{x}$
\be
\label{exp}
b (\X, x, \bar{x} ) = \sum
b(\X)_{a_1 b_1, a_2 b_2, \ldots ;\da_1 \db_1 ,\da_2 \db_2, \ldots }
x^{a_1 b_1} x^{a_2 b_2} \ldots ;
\bar{x}^{\da_1\db_1}\bar{x}^{\da_2\db_2}\ldots
\ee
are totally symmetric both in undotted and dotted indices
(equivalently, these equations are recognized as
complexified $3d$ equations (\ref{oscal}) to be solved by the $3d$
 twistor ansatz). In particular, the holomorphic and antiholomorphic
parts
\bee
c (\X, x ) &=& \sum c(\X)_{a_1 a_2 a_3 a_4\ldots}
x^{a_1 a_2} x^{a_3 a_4} \ldots ;\nn\\
\bar{c} (\X,\bar{x} ) &=& \sum \bar{c}(\X)_{\da_1 \da_2, \da_3 \da_4 \ldots }
\bar{x}^{\da_1\da_2}\bar{x}^{\da_3\da_4}\ldots
\eee
expand in powers of $x$ and $\bar{x}$ with totally symmetric
coefficients $c(\X)_{a_1 \ldots a_{2s}}$,
$\bar{c}(\X)_{\da_1 \ldots \da_{2s}}$ to be identified with the
(anti)selfdual $4d$ components of the higher spin fields. These satisfy
the equations (\ref{Xx}) equivalent to
\be
\label{4hs}
\epsilon^{ac}
\f{\p^2}{\p x^{ab} \p {\X}^{c\da}} c (\X, x )=0\,,\qquad
\epsilon^{\da\dd}
\f{\p^2}{\p \bar{x}^{\da\db} \p {\X}^{c\dd}}
\bar{c} (\X,\bar{x} )=0\,,
\ee
which, in turn, reduce to the usual $4d$ massless equations for
massless fields of all spins $c(\X)_{a_1 \ldots a_{2s}}$ and
$\bar{c}(\X)_{\da_1 \ldots \da_{2s}}$. Let us note that for all spins
$s\neq 0$ the Klein-Gordon equation is a consequence of (\ref{4hs})
by virtue of
\be
\epsilon^{ab} \epsilon^{\da\db}
\f{\p^2}{\p \X^{a \da} \p {\X}^{b\db}}
\f{\p}{\p x^{cd}}
c(\X, x )=0\,,\qquad
\epsilon^{ab} \epsilon^{\da\db}
\f{\p^2}{\p \X^{a \da} \p {\X}^{b\db}}
\f{\p}{\p \bar{x}^{\dc\dd}}
 \bar{c}(\X, \bar{x} )=0\,. \nn
\ee
For the $s=0$ field described by  $c({\cal X}) = c(\X,0) = \bar{c}(\X,0)$
it cannot be derived this way because $\f{\p}{\p x^{cd}} c(\X)=
\f{\p}{\p \bar{x}^{\dc\dd}} \bar{c}(\X)=0$.

Clearly, a number of coordinates
$x$ and $\bar{x}$ in the expansions for $c (\X, x )$ and
$\bar{c}(\X,\bar{x} )$ associated with spin can be equivalently
described by a cyclic variable $\phi$ introduced previously.
The fermionic case can be analyzed analogously.

 The component fields
$c(\X)_{a_1 \ldots a_{2s}}$ and
$\bar{c}(\X)_{\da_1 \ldots \da_{2s}}$  describe, respectively,
self-dual and antiself-dual components of the generalized Weyl
tensors of massless fields of
all spins $s$ (both integer contained in $b(X)$ and half-integer
contained in $f_\ga (X)$). In particular,
$c(\X)_{a_1 a_{2}}$ and
$\bar{c}(\X)_{\da_1  \da_{2}}$  describe (anti) self-dual Maxwell
field strengths,
$c(\X)_{a_1 a_2 a_{3}}$ and
$\bar{c}(\X)_{\da_1 \da_2 \da_{3}}$  describe (anti) self-dual
gravitino field strengths,
$c(\X)_{a_1 \ldots a_{4}}$ and
$\bar{c}(\X)_{\da_1 \ldots \da_{4}}$
describe (anti) self-dual Weyl tensors, etc. Remarkably, the $4d$
electro-magnetic
duality transformation and its extension to all higher spins \cite{Hullem}
\be
\label{em}
c(\X)_{a_1 \ldots a_{2s}} \to \exp[{2s i\varphi }]
c(\X)_{a_1 \ldots a_{2s}}\,,\qquad
\bar{c}(\X)_{\da_1 \ldots \da_{2s}} \to \exp[{-2si\varphi }]
c(\X)_{\da_1 \ldots \da_{2s}}\,,
\ee
acquires a purely geometric origin in the generalized space-time $\M_4$,
being a part of the $SL_4$ generalized
Lorentz transformations with the group element of the form
\be
a^{a}{}_b = \exp[{i\varphi }] \delta^{a}{}_b\,,\qquad
a^{\da}{}_\db = \exp[{-i\varphi }] \delta^{\da}{}_\db\,,\qquad
a^{a}{}_\db =a^{\da}{}_b = 0\,.  \nn
\ee
This transformation belongs to the generalized
$SO(M)$ space rotation because it leaves invariant the time-matrix
$T^{a\db}= \delta^{a\db} $. Moreover, the duality transformation leaves
invariant all space-time coordinates $\X^{a\da}$
of $R\times \gs$. This is why from the perspective of the local Cauchy
surface it acts only on the spin indices. (All other generalized
Lorentz transformations affect the space-time coordinates and therefore
contain derivatives of the $4d$ dynamical fields in the transformation
lows.) Thus, the approach developed in \cite{BHS} and in this paper
incorporates dualities in a natural geometric way  as particular
generalized space-time symmetry transformations.

\section{Quantization}
\label{Quantization}

Although Lagrangian formulation for the
dynamical systems described by the equations (\ref{oscal}) and
(\ref{ofer}) is yet lacking, the form of the general solutions
(\ref{bfo}) and (\ref{ffo}) suggests a natural quantization
prescription. The key property of the general solutions
(\ref{bfo}) and (\ref{ffo}) is that they admit a well-defined
decomposition into positive and negative frequency parts thus
allowing for the  definition of the creation and annihilation
operators $b^+ (\xi )$, $f^+ (\xi )$   and $b^- (\xi )$, $f^- (\xi )$,
respectively.

Let us now give more precise definitions.
To make Gaussian integrals over svector integration variables
 well-defined we introduce complex coordinates
\be
Z^{\ga\gb} = Y^{\ga\gb} +i X^{\ga\gb}\,.  \nn
\ee
The imaginary part of $Z^{\ga\gb}$ is identified with the
coordinates in the generalized space-time $X^{\ga\gb}$.
The real part $Y^{\ga\gb}$
is required to be positive definite. It is
treated as a regulator that makes the Gaussian integrals
well-defined. Physical quantities are obtained in the limit
$Y^{\ga\gb}\to 0$.

The expressions (\ref{bfo}) and (\ref{ffo}) are to be understood
as
\be
\label{cbfo}
b (Z,\bar{Z}) =\f{1}{\pi^{\f{M}{2}}}
\int d^M\xi\, \Big ( b^+ (\xi ) \exp -  \xi_{\ga}\xi_{\gb} \bZ^{\ga\gb}
+b^- (\xi ) \exp -  \xi_{\ga}\xi_{\gb}Z^{\ga\gb} \Big ) \,,
\ee
\be
\label{cffo}
f_\ga (Z,\bar{Z}) =\f{1}{\pi^{\f{M}{2}}}
\int d^M\xi\, \xi_\ga
\Big ( f^+ (\xi ) \exp -  \xi_{\gga}\xi_{\gb} \bZ^{\gga\gb}
+ f^- (\xi ) \exp -  \xi_{\gga}\xi_{\gb}Z^{\gga\gb} \Big ) \,.
\ee
The positive and negative frequency parts identify with the
holomorphic and antiholomorphic parts of the quantum field
\be
b(Z,\bar{Z}) = b^+ (\bZ) + b^- (Z)\,,\qquad
f_\ga (Z,\bar{Z}) = f_\ga^+ (\bZ) + f_\ga^- (Z)\,, \nn
\ee
\be
\label{cbfopm}
b^\pm (Z) =\f{1}{\pi^{\f{M}{2}}}
\int d^M\xi\, b^\pm (\xi ) \exp -  \xi_{\ga}\xi_{\gb} Z^{\ga\gb}\,,
\ee
\be
\label{cffopm}
f^\pm_\gga (Z) =\f{1}{\pi^{\f{M}{2}}}
\int d^M\xi\, \xi_\gga
 f^\pm (\xi ) \exp -  \xi_{\ga}\xi_{\gb} Z^{\ga\gb}\,.
\ee
For the fields $b(Z,\bar{Z})$ and $f_\ga (Z,\bar{Z})$ to be real,
$b^+(\xi)$  and $f^+(\xi)$ have to be complex conjugated to
$b^-(\xi)$  and $f^-(\xi)$, respectively.
The quantum operators $b^\pm (\xi)$
and $f^\pm (\xi)$ are required to have definite oddness according to
(\ref{par}).

We now interpret $b^+(\xi)$, $f^+(\xi)$ and $b^-(\xi)$ , $f^-(\xi)$
as hermitian conjugated bosonic and fermionic
creation and annihilation operators
subject to the commutation relations
\be
\label{qpr}
[b^\pm (\xi_1 ) , b^\pm (\xi_2 )] =0\,,\qquad
[b^- (\xi_1 ) , b^+ (\xi_2 )] =
\half ( \delta (\xi_1 - \xi_2 )+\delta (\xi_1 + \xi_2 ) )\,,
\ee
\be
\label{qprf}
[f^\pm (\xi_1 ) , f^\pm (\xi_2 )]_+ =0\,,\qquad
[f^- (\xi_1 ) , f^+ (\xi_2 )]_+ =
\half ( \delta (\xi_1 - \xi_2 )- \delta (\xi_1 + \xi_2 ) )\,,
\ee
where $[,]_+$ denotes anticommutator.
The vacuum state is defined to satisfy
\be
b^- (\xi ) \rvac =0\,,\qquad
f^- (\xi ) \rvac =0\,, \nn
\ee
\be
\lvac b^+ (\xi ) =0\,,\qquad
\lvac f^+ (\xi )  =0\,. \nn
\ee
\be
\langle 0 \rvac =1\,. \nn
\ee
One-particle states are
\be
\label{1-b}
\int d^M \xi B (\xi) b^+ (\xi)  \rvac \,,\qquad
\int d^M \xi F (\xi) f^+ (\xi)  \rvac  \nn
\ee
with arbitrary complex functions $B(-\xi)=B(\xi )$ and
$ F (-\xi)=-F (\xi)$.

The states are normalizable provided that $B (\xi)$ and
$F (\xi)$ belong to $L^2$
\be
\int d^M \xi\overline{ B (\xi)} B (\xi) < \infty\,,\qquad
\int d^M \xi\overline{ F (\xi)} F (\xi) < \infty\,. \nn
\ee
A useful basis is provided by the functions of the form
\be
B (\xi) = P(\xi ) \exp-T^{\ga\gb}\xi_\ga \xi_\gb  \,,\qquad
F (\xi) = Q(\xi ) \exp-T^{\ga\gb}\xi_\ga \xi_\gb  \,, \nn
\ee
where $P(\xi)$ and $Q(\xi)$ are polynomials of $\xi_\ga$
and $T^{\ga\gb}$ is the positive definite
matrix associated with the time arrow. This basis is equivalent to
the unitary Fock module over the higher spin conformal symmetries,
that was conjectured in \cite{3d,BHS} to be equivalent
to the space of quantum states in the corresponding quantum field
theory. The formulas (\ref{1-b}) thus prove this conjecture.

The quantization prescription (\ref{qpr}) and (\ref{qprf}) allows
us to write the conserved charges associated with the generators of
the $sp(2M)$ transformations (\ref{db}) and (\ref{df})
\be
\label{Pop}
P_{\ga\gb} = \int d\xi^M \Big (
b^+ (\xi ) \xi_\ga \xi_\gb b^- (\xi ) +
f^+ (\xi ) \xi_\ga \xi_\gb f^- (\xi ) \Big )\,,
\ee
\be
\label{Lop}
L_{\gb}{}^{\ga} =-\f{i}{2} \int d\xi^M \Big (
b^+ (\xi ) (\xi_\gb \f{\p}{\p\xi_\ga} + \f{\p}{\p\xi_\ga}\xi_\gb )
b^- (\xi )+
f^+ (\xi )
(\xi_\gb \f{\p}{\p\xi_\ga} +        \f{\p}{\p\xi_\ga}\xi_\gb )
f^- (\xi ) \Big )\,,
\ee
\be
\label{Kop}
K^{\ga\gb} = -\f{1}{4} \int d\xi^M \Big (
b^+ (\xi )
\f{\p^2}{\p \xi_\ga\p \xi_\gb}
b^- (\xi ) +
f^+ (\xi )\f{\p^2 }{\p \xi_\ga\p \xi_\gb}
f^- (\xi ) \Big )\,.
\ee
The supergenerators are
\be
Q_\ga = \int d\xi^M \Big (
b^+ (\xi ) \xi_\ga  f^- (\xi ) +
f^+ (\xi ) \xi_\ga b^- (\xi ) \Big )\,, \nn
\ee
\be
S^\ga = \f{i}{2}\int d\xi^M \Big (
b^+ (\xi ) \f{\p}{\p \xi_\ga}  f^- (\xi ) +
f^+ (\xi ) \f{\p}{\p \xi_\ga}  b^- (\xi ) \Big )\, \nn
\ee
with the obvious anticommutation relations
\be
\{Q_\ga , Q_\gb \}  = 2P_{\ga\gb}\,,\qquad
\{S^\ga , S^\gb \}  = 2K^{\ga\gb}\,,\qquad
\{Q_\ga , S^\gb \}  = - 2L_{\ga}{}^{\gb}\,.\nn
\ee
Note that from these relationships it follows that
averages of the operators $P_{\ga\gb}$ and $K^{\ga\gb}$ form
some positive semi-definite matrices as it is also obvious
from (\ref{Pop}) and (\ref{Kop}). In particular, the energy
operator
\be
E= \f{1}{M} T^{\ga\gb} P_{\ga\gb} \nn
\ee
is positive semi-definite for any positive definite matrix $T^{\ga\gb}$
associated with the chosen time direction. Note also that
bosonic and fermionic
vacuum energies in (\ref{Pop})-(\ref{Kop}) cancel out as a
consequence of supersymmetry.

Let us now define the $\D$ functions as
\be
\label{cD-}
\D^-(Z) = \f{i}{ \pi^M}
\int d^M\xi\,  \exp -  \xi_{\ga}\xi_{\gb} Z^{\ga\gb}\,,
\ee
\be
\D^+(\bZ)=-\D^-(\bZ) =\overline{\D^- (Z)} \,.
\ee
We have
\be
[b^- (Z_1 ) , b^+ (\bZ_2 )] =-i \D^- (Z_1 +\bZ_2 )\,, \nn
\ee
\be
[b (Z_1,\bZ_1 ) , b (Z_2 ,\bZ_2 )] =-i \D (Z_1 +\bZ_2 ,\bZ_1 +Z_2)\,,   \nn
\ee
where
\be
\D (Z,\bZ) = \D^- (Z) +\D^+ (\bZ )=\D^- (Z)-\D^- (\bZ)\,. \nn
\ee
By construction, the functions $\D^- (Z)$, $\D^+ (Z)$ and $\D(Z,\bZ)$
solve the equations of motion (\ref{oscal}). A rotation of the
contour of integration over the variables $\xi_\ga$ in the
complex plane gives the following result
\be
\label{Dsol}
\D^- (Z)\Big |_{Y\to 0} =\f{i}{\pi^{\f{M}{2}}}
\exp-\f{i\pi I_X}{4} \f{1}{\sqrt{|det (Z)|}}\Big |_{Y\to 0}\,.
\ee
Here $I_X$ is the inertia index of the matrix
$X^{\ga\gb}$.
\be
\label{iner}
I_X = n_+ - n_- \,,
\ee
where $n_+$ and $n_-$ are, respectively, the numbers of positive and
negative eigenvalues of $X^{\ga\gb}$. The formula (\ref{Dsol})
is in accordance with the explicit check in section \ref{Propagation} that
${|det (X)|}^{-\half}$ provides a solution of the field equations away
from singularities. The integral representation (\ref{cD-})
provides the precise definition of the regularized expression in the
complex plane.

{}From (\ref{Dsol}) it follows that
\be
\D^+ (Z)\Big |_{Y\to 0} =
\f{1}{i\pi^{\f{M}{2}}}
\exp \f{i\pi I_X}{4} \f{1}{\sqrt{|det (Z)|}}\Big |_{Y\to 0}\,, \nn
\ee
and, therefore,
\be
\label{Dfu}
\D (Z)\Big |_{Y\to 0} =\f{2}{\pi^{\f{M}{2}}}
\sin(\f{\pi I_X}{4}  )\f{1}{\sqrt{|det (Z)|}}\Big |_{Y\to 0}\,.\nn
\ee

Extension of the suggested quantization scheme  to the fermionic
case of svector field $f_\ga (X)$ is straightforward
\be
\label{quant2s}
\{ f_\ga^- (Z_1 ), f_\gb^+ (\bZ_2 ) \} = -i \D_{\ga\gb}^- (Z_1+\bZ_2 )\,,
\ee
\be
\label{quant3s}
\{ f_\ga (Z_1,\bZ_1 ), f_\gb (Z_2,\bZ_2 ) \} = -i
\D_{\ga\gb} (Z_1+\bZ_2, \bZ_1 +Z_2 )\,,
\ee
where
\be
\D_{\ga\gb} (Z,\bZ)=\D_{\ga\gb}^- (Z) + \D_{\ga\gb}^+ (\bZ) \,,\qquad
\D_{\ga\gb}^+ (Z) =\D_{\ga\gb}^- (Z)\,, \nn
\ee
\be
\D^-_{\ga\gb} (Z) = -\f{\p}{\p Z^{\ga\gb}} \D^- (Z)\,. \nn
\ee

{}From (\ref{Dfu}) it follows that bosonic and fermionic
$\D -$functions vanish if
$I_X = 4n$, being different from zero otherwise.  Since the
indices $\ga$ and $ \gb$ take even number of values $M$,
$I_X = n_+ - n_- = M - 2n_- $ is even. Therefore
\be
\label{D0}
\D (Z,\bZ) \neq 0\qquad \mbox{for} \quad I_X = 4n+2\quad n\in {\bf Z}\,.
\ee

For the case of $M=2$ corresponding to the usual $3d$ geometry,
this leads to the standard picture that the $\D-$function is zero
for the space-like separation $X$ with $I_X =0$ and is different from
zero for the future ($I_X =2$) and past ($I_X =-2$) cones. For
the first sight, the situation for $M>2$ looks unsatisfactory because
the $\D$ functions can be different from zero for the space-like
separations with $I_X \neq \pm M$. The point
however is that there is no reason to require the $\D -$function to
vanish on the global Cauchy surface $\Sigma$ as a whole,
where the fields cannot be
localized. The microcausality requires instead the $\D -$function
to have usual properties upon restriction to the
space-time $R^1\times \gs$  with some local
Cauchy surface $\sigma$ as a space manifold.
In particular, this is the case
 for the $4d$ coordinates $\X^{a\da}$ of
(\ref{4dfib}) at $x^{ab} = \bar{x}^{\da\db} =0$ because the (real)
inertia index of matrices of this form
can only take values $+4$ (future), $0$ (space-like
separation) or $-4$ (past). As a result the corresponding $4d$
$\D-$ function vanishes inside the future and past cones. It is however
different from zero at the boundaries of the future and past cones  which
conclusion respects the microcausality principle
and is in accordance with the properties of
the usual $\D -$function for $4d$ massless fields
known to be localized at the boundary of the cone
(see e.g., \cite{BSh}). The $4d$ $\D-$function is a
distribution localized at zeros of the eigenvalues of $X^{\ga\gb}$.
In section \ref{Towards any M} we argue that the microcausality for
higher $M$ is also
respected provided the local Cauchy surface $\gs$
is associated with an appropriate
Clifford algebra in which case $I_X$ can only take
three values associated with the past, future and space-like separations.

Note that choosing the opposite type of the commutation relations,
would  replace $\sin(\f{\pi I_X}{4}  )$ in
(\ref{Dfu}) by $\cos(\f{\pi I_X}{4}  )$. As a result,
the corresponding $\D -$function would be different from zero for
the space-like arguments with $I_X =0$ thus violating
the microcausality principle. Therefore, analogously to
the case of Minkowski space-time, the locality requirement fixes
statistics of the fields in the usual way requiring the scalar and
svector fields to obey the Bose and Fermi statistics, respectively.

Let  $\Theta (X)$ be the characteristic function of the
future cone, i.e.
\bee
\Theta (X) &=& 1\qquad \mbox{$X$ is positive semi-definite,}\nn\\
\Theta (X) &= &0\qquad \mbox{otherwise}\,.
\eee
The advanced, retarded and causal Green functions are defined as
\be
G^{ret}(X)= \Theta (X) \D (Z,\bZ)\Big |_{Y\to 0}\,,\qquad
G^{adv}(X)= -\Theta (-X) \D (Z,\bZ)\Big |_{Y\to 0}\,, \nn
\ee
\be
G^{c} (X) =  \Big (\Theta (X) \D^-(Z )
 -\Theta (-X) \D^+ (\bZ) \Big )\Big |_{Y\to 0}\,. \nn
\ee

\section{Towards any M}
\label{Towards any M}

In the general case, the local Cauchy bundle $E$
is $M$-dimensional. Although a
full analysis of the structure of $E$ is beyond the
scope of this paper, as a first step towards the general case
we wish to emphasize the role of the Clifford algebras in the
generalized space-time geometry.

In the examples of $M=2$ and $M=4$ the space coordinates were
associated with the set of symmetric matrices $\sigma^{\ga\gb}_n$
\be
\label{xsig}
x^{\ga\gb} = x^n \sigma_n^{\ga\gb}
\ee
such that the matrices
\be
\label{ggs}
\gamma_i{}^\ga{}_\gb = \sigma_i{}^{\ga\gga}T_{\gga \gb}\qquad i=1\ldots d-1
\ee
satisfy  the Clifford algebra relationships
\be
\label{clif}
\gamma_i{}^\ga{}_\gga \gamma_j{}^\gga{}_\gb +
\gamma_j{}^\ga{}_\gga \gamma_i{}^\gga{}_\gb =2\eta_{ij}\delta^\ga_\gb\,,
\ee
where $\eta_{ij}$ is some positive definite symmetric
form (for example, one can choose a basis with $\eta_{ij} = \delta_{ij}$).
There are several reasons  why the
base space manifold $\sigma$ is likely to be associated with the
Clifford algebras for the general case.

An immediate consequence of
(\ref{clif}) is  that the matrices $\gs_n^{\ga\gb}$ are traceless
\be
\label{gst}
\gs_n^{\ga\gb} T_{\ga\gb} =0\,,
\ee
whenever $d\geq 3$, thus belonging to the global Cauchy surface.
Another important property is that the momenta
\be
\label{mom}
k_n (\xi ) = \sigma_n^{\ga\gb} \xi_\ga\xi_\gb\,
\ee
map the cone $R^M /Z_2$ on $R^{d-1}$, i.e. varying real twistor parameters
$\xi_\ga$ it is possible to get arbitrary values of $k_n (\xi ) $. This
results from the invariance of the construction under the space rotations
$SO(d-1)$ generated by
\be
\label{srot}
M_{nm}= \f{1}{4}[\gamma_n , \gamma_m ]\,.
\ee
By a space rotation one aligns a vector $k_n (\xi )$ along any
direction and then normalizes it arbitrarily by a rescaling of $\xi_\ga$.
That momenta $k_n (\xi )$
span $R^{d-1}$ allows for localization of the fields in $d-1$ space
$x^n$ coordinates dual to $k_n$, i.e., by means of integration over
$k_n$, one can reach the delta-functional initial data
$\delta (x^n - x^n_0 )$ localized at any point of the physical space
$R^{d-1}$.

Because the square of any linear combination of
 $\gamma$ matrices is proportional to the unit matrix, for any vector
$a^n$ there exists such a basis in the space of $\xi_\ga$ that
\be
\label{keyrep}
T^{\ga\gb}= \delta^{\ga\gb}\,,\qquad a^n \sigma_n^{\ga\gb} =
\sqrt{a^2} Y^{\ga\gb}\,,\qquad a^2 =a^n a^m \eta_{nm}\,,
\ee
where
\be
Y= \left( \begin{array}{cc}
                   I    &   0   \\
                   0    &  -I
            \end{array}
    \right)  \nn
\ee
with all four blocks being $\f{M}{2} \times \f{M}{2} $ matrices
($M$ is assumed to be even). As a result, the
(non-degenerate) space-time matrix coordinates of the form
\be
\label{tX}
X^{\ga\gb} = t T^{\ga\gb} + x^n \sigma_n^{\ga\gb}
\ee
can only have three values of the inertia index (\ref{iner})
\bee
\label{clcon}
I_X &=& M\qquad \mbox{for}\quad t>\sqrt{x^2}\,,\nn\\
I_X &=& - M\qquad \mbox{for}\quad t< - \sqrt{x^2}\,,\nn\\
I_X &=& 0\qquad \mbox{for}\quad t^2 < {x^2}\,,
\eee
where
\be
x^2 =x^n x^m \eta_{nm} \,.  \nn
\ee
This corresponds to the standard space-time picture with the
future cone ($I_X >0$), past cone ($I_X <0$) and the space-like
region ($I_X =0$). In accordance with the consideration of section
\ref{Quantization} this property of space Clifford coordinates
implies microcausality in the space-time $R\times \gs$.

Note that the condition that $I_X$ can take
in the linear space of matrices of the form (\ref{tX})
only maximal value $M$ (future), minimal value $-M$
(past) or some fixed  intermediate value
$I_X =I_{space}\neq \pm M$ for all values of $t$ and $x^n$
can be taken as an alternative definition leading to
the Clifford algebra relations (\ref{clif}).
Actually, it is true if for any
$x^n$ the matrix
$ x^n \sigma_n^{\ga\gb} T_{\gb\gga}$ has just two different eigenvalues.
Since the sign change of $x^n$ maps $I_{space} \to -I_{space}$,
the only consistent choice is $I_{space}=0$. Assuming that the
time $t$ is defined so that the
matrices $\sigma_n^{\ga\gb}$ associated with the space coordinates
are $T$-traceless this implies that the matrix
$ x^n \sigma_n^{\ga\gb}T_{\gb\gga}$ has $\f{M}{2}$ eigenvalues
$\mu$ and $\f{M}{2}$ eigenvalues $-\mu$. Equivalent statement is that,
for any $x^n$, the matrix $ x^n \sigma_n^{\ga\gb}T_{\gb\gga}$ is
traceless and its matrix square is proportional to the unit matrix
that is equivalent to the Clifford algebra  definition (\ref{clif}).

These properties indicate that the Clifford algebra realization of
space is closely related to the concept of locality and microcausality.
In other words, the generalized space-time $\M_M$ is visualized
via  Clifford algebras. Let us note that the space metric
$\eta_{nm}$ appears in the theory just by identification of an
appropriate Clifford algebra (\ref{clif}).

The Clifford realization of the space-time  $R\times \gs$
guarantees usual conformal symmetry in $d$ dimensions.
The ordinary space rotation symmetry $o(d-1)$ is generated
in the standard way as the subalgebra of the generalized Lorentz
symmetry $sl_M$ spanned by the generators (\ref{srot}). Its extension
to the Lorentz subalgebra $o(d-1,1)\subset sl_M$ is achieved by
boosts realized as
\be
\label{boo}
l_n = \gga_n\,.
\ee
The Lorentz algebra extends to the Poincare algebra by the transformations
(\ref{db}), (\ref{df}) with the parameters
\be
\gge^{\ga\gb} = a^0 T^{\ga\gb} + a^k \gs_k^{\ga\gb}\,. \nn
\ee
Its further extension to the conformal algebra is achieved via the
generalized special conformal transformations
(\ref{db}), (\ref{df}) with the parameters
\be
\gge_{\ga\gb} = b^0 T_{\ga\gb} + b^k \gs_k{}_{\ga\gb}\,. \nn
\ee
Recall that the dilatation appears as the central component in the
$gl_M$ extension of the generalized Lorentz transformations, i.e.
it is generated by the unit element of the Clifford algebra.

The (classical) generalized electro-magnetic duality group
identifies with such a
subgroup of the generalized Lorentz transformations $SL_M$
that leaves invariant the time matrix $T^{\ga\gb}$ and the space
coordinates of the local Cauchy surface. By definition, such defined
duality group acts on the fiber of the local Cauchy bundle $E$, that is
on the indices of the usual space-time fields in $R\times \gs$.

For $M=2^p$ the algebra of real matrices $Mat_{2^p}$ is isomorphic
to a particular real Clifford algebra. Despite the system was shown
to be Lorentz covariant, this does not necessarily mean that the set
of space $\gga_n$ matrices associated with the local Cauchy surface
admits an extension by a matrix $\gamma_0$
\be
\label{g0}
\gamma_0{}^\ga{}_\gga \gga_0{}^\gga{}_\gb = -\delta^\ga_\gb\,
\ee
that anticommutes to the space-like matrices $\gamma_i$.
Leaving details for a future publication \cite{pr}, let us just mention
that $\gga_0$ exists when the Clifford algebra has
antisymmetric charge conjugation
matrix ($p = 1$ or $2$ $mod\quad 4$) but does not exist otherwise
($p = 0$ or $3$ $mod\quad 4$). This fact has an important interpretation.
Namely, the cases that do not allow Lorentz invariant extension of the
Clifford algebra are chiral, i.e. svector indices correspond to
left or right real spinors (depending on the definition of $\gga$ in view
of the automorphism $\gga \to -\gga$). Indeed, when $\gga_0$ exists,
the boost generators can be identified as usual with $\gga_k \gga_0$.
The operator $\Gamma = \gga_0 \gga_1 \ldots \gga_{d-1}$ then allows to
define chirality in the standard way, thus giving rise to left and right
svectors (may be complex conjugated to each other). If $\gga_0$ does not
exist one has to use the realization (\ref{boo}) for boosts implying
that the svector representation forms an irreducible (and, therefore,
chiral) representation of the Lorentz algebra.
As a result, the corresponding theory
as a whole turns out to be chiral, describing irreducible (anti)self-dual
conformal fields in $d$ dimensions. Recall that
conformal fields are described by scalar, spinor and, for $d$ even, by
massless fields associated with the representations of the
little group $o(d-2)$ described by rectangular Young diagrams of
maximal height $\half(d-2)$ \cite{sieg,met}.

Let us consider the important example of $M=8$. The corresponding real
Clifford algebra is defined by the relationships
\be
\{ \psi_A \,,\psi_B \} = - \delta_{AB}\,,\qquad A,B = 1-6\,. \nn
\ee
The charge conjugation matrix $C_{\ga\gb}=C_{\gb\ga}$ is symmetric and
positive definite. It can therefore be identified with the time-defining
matrix $T_{\ga\gb}$.
The matrices
\be
\label{gm}
\psi_{A_1 \ldots A_n\,\ga\gb}= \psi_{ [A_1}\ldots \psi_{A_n ]}{}_{\ga\gb}\,
\ee
turn out to be symmetric in $\ga ,\gb$  for $n= 0,3$ and  $4$ and
antisymmetric for $n= 1,2,5,6$ in an irreducible representation.
Here the indices $\ga$ and $\gb$ are raised and lowered by the
charge conjugation matrix $C_{\ga\gb}$ identified with
$\psi_{A_1 \ldots A_n\,\ga\gb}$ at $n=0$.
Being antisymmetric, the matrices $\psi_A$ cannot serve themselves as
a basis for coordinates of a local Cauchy surface in $\M_M$. One can
however choose five symmetric matrices
\be
\label{gga6}
\gga_n  = \psi_{n56} \quad for\quad n= 1\ldots 4\,,\qquad
\gga_n  = \psi_{1234} \quad for \quad n= 5\,
\ee
that satisfy the  Clifford algebra relations
\be
\{ \gga_n \,,\gga_m \} = 2\delta_{nm}\,,\qquad n,m = 1\ldots 5\,. \nn
\ee
These matrices are traceless and can therefore be identified with a
particular basis of space coordinates. Along with the time matrix
$C_{\ga\gb} = T_{\ga\gb}$ this defines a $6d$ space-time. The
matrices (\ref{gm}) do not contain a matrix $\gga_0$ that anticommutes
to $\gga_n$. Eight-component svector therefore identifies with a
chiral $6d$ spinor. The properties of the $\M_8$ model make it
reminiscent of the $6d$ (super)conformal theory proposed
by Hull \cite{Hull}. We expect that the $M=8$ theory is a higher spin extension
of the $6d$ conformal self-dual gravity theory studied by Hull \cite{Hull}.
For supersymmetrization of the model for any $M$ see
\cite{BHS} and section \ref{Extended Supersymmetry}
of this paper. In \cite{BHS} it was also explained
for the example of $4d$ (i.e. $M=4$) theory how
one can truncate the model to a particular (lower spin, if desirable)
irreducible supermultiplet by virtue of certain auxiliary
noncommutative scalar fields.

Let us note that, excluding the fifth coordinate associated with $\gga_5$
from the set of space coordinates allows one to introduce the $5d$
time-like matrix $\gga_0$ by identifying it with $\psi_{123456}$.
In fact, one can treat such a model as a (non-chiral)
result of compactification of the original $6d$ chiral model on $S^1$
upon an appropriate identification  in the momentum space.

The matrices $\gga_n$ provide a basis for the coordinates of the $M=8$
local Cauchy surface $R^5$. Let us now show that the rest three coordinates
of the local Cauchy bundle are associated with the group manifold $SO(3)$,
i.e. $E=R^5 \times SO(3)$
\footnote{Note that the idea that
the $M=8$ twistor space identifies with $R^5\times S^3$ was also
suggested in the context of the analysis of the world-particle models
by Lukierski \cite{Luk}.}. The point is that the five generalized space
momenta
\be
k_n (\xi ) = \gga_{n\ga\gb}\xi^\ga\xi^\gb \nn
\ee
are invariant under the $SU(2)$ rotations of $\xi_\ga$ generated by
$\psi_{5} , \psi_6$ and $\psi_{56}$. The coordinates of $SU(2)$ are
analogous to the cyclic coordinate $\phi$ in the $M=4$ case. Because we
are only interested in bilinear combinations in the twistor momenta
$\xi_\ga$ the $SU(2)$ reduces to $SO(3)$. Analogously to the $4d$ case,
the modes on $SO(3)$ are expected to be associated with various $6d$
(generalized) higher spin massless fields. Moreover,  $SO(3)$
is expected to be a (classical) electro-magnetic duality group of the
$M=8$ model because, by construction, it is a subgroup of the generalized
Lorentz transformations that leaves invariant the time coordinate and the
coordinates of the local Cauchy surface $\gs$. It would be interesting
to compare its component action with the $6d$ higher spin duality
transformations discussed by Hull \cite{Hullem}.

For the $M=16$ case with generalized conformal symmetry $Sp(32)$ the
situation is analogous to $M=8$. The Clifford basis elements (\ref{gm})
are symmetric for $n=0,3,4,7,8$ and antisymmetric for $n=1,2,5,6$. The
maximal number of 9 Clifford space coordinates can be identified with
$\psi_{A_1\ldots A_n\,}{}_{\ga\gb}$ ($A=1\ldots 8$) at $n=7$ and $n=8$.
This implies a nine-dimensional local Cauchy surface $\gs$ and, therefore,
ten-dimensional space-time. Again, the corresponding $10d$ theory is
chiral because the corresponding Clifford algebra associated with the
space coordinates does not admit an extension to the
$10d$ Minkowski case. It becomes a non-chiral
relativistic $9d$ theory upon
compactification of one of the space dimensions. The $M=16$
local Cauchy  bundle is expected to have a seven-dimensional fiber.
The world line analysis of the twistor dynamics  suggests
\cite{Luk} that the relevant choice of the $M=16$ Cauchy bundle may be
$E = \gs \times S^7$, $\gs = R^9$.

The $M=32$ model possessing the generalized conformal symmetry
$Sp(64)$ is a relativistic $11d$ theory and, as such, can be related
to $M$ theory. (Note that the relevance of 64 supercharges
 to $M$-theory was discussed in \cite{Bars}.)

A detailed analysis of $M>4$ models requires some technicalities on
the Clifford algebra realization of $\M_M$ and associated
symmetries and will be given elsewhere \cite{pr}.
One hard issue is to analyze higher $M$ analogues of the changes of
variables (\ref{k2}) and (\ref{k3}), (\ref{cyc}).

\section{Extended Supersymmetry}
\label{Extended Supersymmetry}

The unfolded form of the superfield extended supersymmetry generalization of
the equations (\ref{oscal}) and (\ref{ofer}) was discussed in \cite{BHS}.
Here we would like to discuss a slightly different, although equivalent,
formulation of an extended supersymmetric system that exhibits the
supersymmetry
$osp(2L,2M)$. Namely, let us introduce the generalized supercoordinates
$X^{AB}$ with $A= (\ga, i)$, $\ga=1\ldots M$, $i=1\ldots L$. Let $\pi(A)=0$ for
$A=\ga$ and $\pi (A) =1$ for $A=i$. The coordinates $X^{AB}$ are required to be
graded symmetric
\be
X^{AB}= (-)^{\pi (A) \pi (B)} X^{BA}\,.  \nn
\ee
The coordinates
$X^{\ga\gb}=X^{\gb\ga}$ and $X^{ij}=-X^{ji}$ are even (commuting) while $X^{\ga
i}=X^{i \ga}$ are odd (i.e., anticommuting elements of Grassmann algebra).
Note that the anticommuting supercoordinates  can be identified with a half of
the superspace coordinates $\theta^{\ga i}$ of  section 7.3 of \cite{BHS}.
The formulation presented here can be thought of as a sort of
a chiral superfield formulation compared to that of \cite{BHS}.
The coordinates $X^{ij}$ are new.

The straightforward generalization of the equation (\ref{oscal}) is
\be
\label{soscal}
\f{\p^2}{\p X^{AB} \p X^{CD}} \Phi (X) = (-1)^{\pi(C)\pi(B) }
\f{\p^2}{\p X^{AC} \p X^{BD}} \Phi (X)\,.
\ee
This equation is invariant under the
straightforward extension of the $sp(2M)$ transformations
(\ref{PS})-(\ref{PK}) to $osp(2L, 2M)$ with appropriate grading-dependent
signs inserted. Generic solution of  (\ref{soscal}) is analogous to
(\ref{bfo})
\be
\label{sbfo}
\Phi (X) =\f{1}{\pi^{\f{M}{2}}}
\int d^{M,L}\eta\, \Big ( \Phi^+ (\eta ) \exp i  \eta_{A}\eta_{B} X^{AB}
+\Phi^- (\eta ) \exp -i  \eta_{A}\eta_{B}X^{AB} \Big ) \,,
\ee
where $\eta_A = (\xi_\ga, \psi_i )$ with $\xi_\ga$ and $\psi_i$ being,
respectively, even and odd $(\psi_i \psi_j = - \psi_j \psi_i )$
integration variables. We require
\be
\Phi^\pm (-\eta ) =(-1)^{M+L}\Phi^\pm (\eta )\, \nn
\ee
(relaxing the condition that $M$ is even).
As a result, the $osp(2L, 2M)$ invariant system (\ref{soscal}) turns
out to be equivalent to the set of the fields $b(\xi,\psi)$ and
$f(\xi,\psi )$ in $\M_M$ satisfying
$b(\xi, -\psi ) = (-1)^{L+M}b(\xi, \psi )  $,
$f(\xi, -\psi ) = -(-1)^{L+M}f(\xi, \psi )  $. In other words,
a single field $\Phi (X)$ in the generalized superspace
contains a set of bosons and fermions described by all antisymmetric
tensors of even (odd) and odd (even) ranks, respectively, for $M+L$ even
(odd). Note that the generators of $osp(2L,2M)$ can be realized as quantum
operators analogous to (\ref{Pop})-(\ref{Kop}) associated with various
bilinear combinations of $\eta_A$ and $\f{\p}{\p \eta_A}$.

In fact, the relevance of the equations (\ref{soscal}) is most obvious
from their unfolded form
\be
\label{unfs}
(\f{\p}{\p X^{AB}}  -\f{\p^2}{\p Y^A \p Y^B}) \Phi (Y|X) =0\,,\qquad
\Phi (-Y|X) = \Phi (Y|X)\,, \nn
\ee
where $Y^A = (y^\ga ,\go^i )$ are auxiliary supercoordinates.
Following to the methods of unfolded dynamics (see \cite{BHS,Gol} and
references therein) it is elementary to see that the system (\ref{unfs})
is $osp(2L,2M)$ invariant, is equivalent to  (\ref{soscal})
for $\Phi (X)=\Phi (0|X)$ and reduces to the system of bosons and
fermions in $\M_M$ associated, respectively, with odd and even elements
of the Grassmann algebra generated by $\go^i$, dual to
the Grassmann algebra generated by $\psi_i$.

\section{Geometric Origin of the Generalized Space-Time}
\label{CM}

The group $Sp(2M)$ is constituted by the real matrices
\be A= \left( \begin{array}{cc}
                   a    &   b   \\
                   c   &   d
            \end{array}
    \right)
{}
\label{1}
\ee
with $M\times M$ blocks  $a_\ga{}^\gb$, $b_{\ga\gb}$, $c^{\ga\gb}$
and $d^\ga{}_\gb$ satisfying relations
\be
\label{Sp1}
a_\ga{}^\gga b_{\gb\gga}=a_\gb{}^\gga b_{\ga\gga}\,,
\ee
\be
\label{Sp2}
c^\ga{}_\gga d^{\gb\gga}= c^\gb{}_\gga d^{\ga\gga}\,,
\ee
\be
\label{Sp3}
a_\ga{}^\gga d^{\gb}{}_{\gga}- b_{\ga\gga} c^{\gb\gga} =\delta_\ga^\gb\,
\ee
equivalent to the invariance condition $ACA^t = C$
for the skewsymmetric bilinear form
\be
C= \left( \begin{array}{cc}
                   0    &   I   \\
                   -I   &   0
            \end{array}
    \right)\,,
\label{2}
\ee
where $I$ is the $M\times M$ unit matrix.

$Sp(2M)$ contains the subgroup $T$ constituted by the elements
\be
t(X)= \left( \begin{array}{cc}
                   I    &   X   \\
                   0    &   I
            \end{array}
    \right)\,
\label{4}
\ee
with various real generalized coordinates $X^{\ga\gb}$. The
group of translations $T$ is Abelian and has the product low
\be
t(X)t(Y) = t(X+Y)\,. \nn
\ee
The subgroup of generalized Lorentz transformations and
dilatations is described by
the matrices (\ref{1}) with $b=c=0$ and
$a_\ga{}^\gga d^{\gb}{}_{\gga} =\delta_\ga^\gb\,.$ The subgroup of
special conformal transformations is constituted by the matrices
(\ref{1}) with $a=d=I$, $b=0$.

Let $P_M$ be the parabolic subgroup of $Sp(2M)$ constituted by the
matrices (\ref{1})-(\ref{Sp3}) with $b^{\ga\gb}=0$, i.e.,
\be P_M\ni p= \left( \begin{array}{cc}
                   a    &   0  \\
                   c   &   d
            \end{array}
    \right)\,.
\label{p}
\ee

The compactified generalized space-time is the coset space
\be
\label{DC}
\C\M_M = Sp(2M)/P_M  \nn
\ee
constituted by the elements $h\in Sp(2M)$ identified modulo the
right action of $P_M$
\be
h\sim h_1 = h p\,,\qquad h\in Sp(2M)\,,\quad p\in P_M\,.
\ee
$\C\M_M$ consists of the classes represented by elements
$t(X)$ of the group of translations $T$, which identify with the
points of the uncompactified generalized space-time $\M_M$, along
with some additional equivalence classes that represent conformal infinity.

Any $Sp(2M)$ group element $A$ (\ref{1}) with a nondegenerate
block $d$ is in the class represented by some $t(X)\in T$. Indeed,
once $det \Big |d^\ga{}_\gb\Big | \neq 0$, $A=A^\prime C^\prime$ with some
\be
A^\prime= \left( \begin{array}{cc}
                   a^\prime    &   b^\prime   \\
                   0    &   d^\prime
            \end{array}
    \right)\,,\qquad
C^\prime= \left( \begin{array}{cc}
                   I    &   0   \\
                   c^\prime    &   I
            \end{array}
    \right)\, \nn
\ee
and then $A^\prime = t(X)\tilde{A}$ where $\tilde{A}$ has only
diagonal blocks nonzero. As a result, any element of $Sp(2M)$
with $det \Big |d^\ga{}_\gb\Big | \neq 0$
belongs to some equivalence class associated with the
uncompactified generalized space-time $\M_M$.

{}From (\ref{Sp3}) it follows that $d$ is non-degenerate for
any element $p$ (\ref{p}) of the parabolic subalgebra $P_M$. As a result,
$rank |d|$ of an element $A\in Sp(2M)$
(\ref{1}) is the same for all $Ap$, $p\in P_M$. In other words,
$rank |d|$ characterizes different types of the equivalence classes, i.e.
different subsets of the compactified space-time $\C\M_M$. The subset of
elements with $det|d|= M$ identifies with $\M_M$. Those with
$rank |d| = m$, $m= 0,1,2, \ldots M-1$ describe the conformal infinity
strata mentioned in section \ref{Introduction}.

The inversion $R$ is now a well-defined transformation
in $\C\M_M$. Consider the following element of $Sp(2M)$
\be
\tilde{R}= \left( \begin{array}{cc}
                   0    &   I   \\
                   -I    &   0
            \end{array}
    \right)\,,
\label{R}
\ee
i.e. $\tilde{R}=A$ (\ref{1}) with $a_\ga{}^\gb=0$,  $d^\ga{}_\gb =0$,
$b_{\ga\gb} = \delta_{\ga\gb}$, $c^{\ga\gb}=-\delta^{\ga\gb}$
(note that the conditions (\ref{Sp1})-(\ref{Sp3}) are satisfied).
It follows that
\be
\tilde{R}\, t(X)= \left( \begin{array}{cc}
                   0    &   I   \\
                   -I    &   -X
            \end{array}
    \right)\,.\nn
\ee
Choose $p\in P_M$ in the form
\be p(X)= \left( \begin{array}{cc}
                   -X    &   0   \\
                   I    &   -X^{-1}
            \end{array}
    \right)\,.\nn
\ee
Then
\be
\tilde{R}\, t(X)\, p(X)= \left( \begin{array}{cc}
                   I    &   -X^{-1}   \\
                   0    &   I
            \end{array}
    \right)\,.
\label{RX}
\ee
Thus $\tilde{R}$ maps a non-degenerate $X$ to
$-X^{-1}$. Up to a sign, this is the inversion (\ref{inv}).
If $X$ is degenerate, $\tilde{R}\, t (X)$ is also well-defined in
$\C\M_M$, mapping $X$ to some element of the conformal infinity
classes.

More generally, it is easy to see that according to the definition (\ref{DC}),
the action of a general element (\ref{1}) in $\M$ is described by the
matrix fraction-linear transformation
\be
\label{fl}
A (X) = (aX +b) (cX+d)^{-1}
\ee
for nondegenerate $(cX+d)$. This formula for the action of $Sp(2M)$ on
the space of symmetric matrices was used in particular in
\cite{H,F}. It reproduces (\ref{Tgl}), (\ref{Lgl}), (\ref{Kgl}) and
(\ref{RX}) as particular cases.

Note that the  minus sign in
the transformation low (\ref{RX}) is not occasional. The group
$Sp(2M)$ does not contain the $PT$ reflection $X^{\ga\gb} \to - X^{\ga\gb}$.
For $PT$ reflection to be included, $Sp(2M)$ has to be extended to
$Sp(2M)\times Z_2$ which can be defined as the group that
leaves the form $C$ invariant up to
a sign. This is equivalent to replacing (\ref{Sp3}) by
\be
\label{ZSp3}
a_\ga{}^\gga d^{\gb}{}_{\gga}- b_{\ga\gga} c^{\gb\gga} =\pm\delta_\ga^\gb\,.
\ee
Simultaneously, $P_M$ has to be extended to $P_M \times Z_2$
with $a_\ga{}^\gga d^{\gb}{}_{\gga} =\pm\delta_\ga^\gb\,$.
The $PT$ reflection is represented by
\be
PT= \left( \begin{array}{cc}
                   I    &   0   \\
                   0    &   -I
            \end{array}
    \right)\,.\nn
\ee
It acts properly on the coordinates $X^{\ga\gb}$ because
$PT\,t(X)\,PT = t(-X)$.
The true inversion (\ref{inv}) $R\in {Sp}(2M)\times Z_2$ is then
represented by
\be
{R}\,= \left( \begin{array}{ccc}
                   0    &   I   \\
                   I    &   0
            \end{array}
    \right)\,.\nn
\ee

The central element of $Sp(2M)$
\be
{F}\,= \left( \begin{array}{ccc}
                   -I    &   0   \\
                   0    &   -I
            \end{array}
    \right)\,\nn
\ee
acts trivially in $\C\M_M$. However, according to (\ref{lorb}) it
acts nontrivially on the sections of the corresponding fiber bundels
over $\C\M_M$ if fermions are present carrying odd numbers of the
svector indices. In other words, $F$ is the boson-fermion
parity operator.

Let us note that alternative definitions of $Sp(2M)-$invariant
$\half M(M+1)-$dimensional matrix spaces were given in \cite{H,F}.
Although we have not check this
explicitly, all three constructions are expected to be equivalent.

The generalization to superspace is straightforward. $Sp(2M)$
is extended to $OSp (L,2M)$
constituted by the supergroup elements
\be
{A}\,= \left( \begin{array}{ccc}
                   a    &   b   & e   \\
                   c    &   d   & f \\
                   g    &   h   & p \\
            \end{array}
    \right)\,
\label{sg}
\ee
that leave invariant the (super)antisymmetric bilinear form
\be
C= \left( \begin{array}{ccc}
                   0    &   I  & 0 \\
                   -I   &   0  & 0 \\
                   0   &   0  &  I \\
            \end{array}
    \right)\,.
\label{sc1}
\ee
The first, second and third rows (columns) in these formulas
have, respectively, heights (widths) $M$, $M$ and $L$. The
superspace  with supercoordinates $X^{\ga\gb}$,
$\theta^{\ga}_i$, introduced in \cite{BHS} corresponds to the
coset space $OSp (L,2M)/P_{L,2M}$ where the parabolic supergroup
$P_{L,2M}$ is formed be the supergroup elements $A\in OSp (L, 2M)$
(\ref{sg}) with $b=0$ and $e=0$.

The superspace of section
\ref{Extended Supersymmetry} of this paper results from  the
decomposition of $OSp(2L,2M )$
\be
{A}\,= \left( \begin{array}{cccc}
                   a    &   b   & e  & p \\
                   c    &   d   & f & q \\
                   g    &   h   & n & r  \\
                   v   &   u   & m & l  \\
            \end{array}
    \right)\,
\label{sg2}
\ee
with the  invariant supersymplectic form
\be
C= \left( \begin{array}{cccc}
                   0    &   I  & 0 &0\\
                   -I   &   0  & 0 &0\\
                   0   &   0  &  0 &I\\
                   0   &   0  &  I &0\\
            \end{array}
    \right)\,
\label{sc2}
\ee
and the parabolic supergroup $P^\prime_{2L,2M}$ formed by the
elements $A\in OSp(2L,2M)$ of the form
\be
{A}\,= \left( \begin{array}{cccc}
                   a    &   0   & e  & 0 \\
                   c    &   d   & f & q \\
                   g    &   0   & n & 0  \\
                   v   &   u   & m & l  \\
            \end{array}
    \right)\,.
\label{p2}
\ee

\section{Outlook}
\label{Conclusions}

It is shown that the equations of motion
(\ref{oscal}) and (\ref{ofer}) in the generalized
space-time proposed in \cite{BHS}
admit consistent interpretation compatible with causality both
at the classical and quantum levels. The coordinates of the
generalized space-time $\M_M$ are various symmetric
$M\times M$ matrices $X^{\ga\gb}$, $\ga,\gb = 1\ldots M$. The
future and past cones of the origin of coordinates
$X^{\ga\gb} =0$ identify with the positive-definite and
negative-definite matrices $X^{\ga\gb}$. The generalized
space-time is shown to have only one
time coordinate associated with any
positive-definite matrix $T^{\ga\gb}$ (or positive semi-definite
for light-like directions). Different choices of
$T^{\ga\gb}$ correspond to different coordinate frames related by
generalized Lorentz transformations.

A global Cauchy surface $\Sigma$ is defined as such a submanifold
of $\M_M$ that any two
its points are separated by a space-like interval and the set of
points that belong to the future and past cones  of all
points of $\Sigma$ covers the whole generalized space-time.
A particular realization of $\Sigma$ is provided by
matrices $X^{\ga\gb}$ satisfying
$T_{\ga\gb} X^{\ga\gb} = 0,$ where
$ T_{\ga\gb} $ is the inverse of  $T^{\ga\gb}$.

A local Cauchy bundle $E$ is a $M-$dimensional space
that provides the full set of unrestricted initial data for the problem.
The base space $\gs$ of $E$, called local Cauchy surface,
is identified with the usual $d-1$ dimensional space.
The difference between the concepts of global Cauchy surface and
local  Cauchy bundle is due to the fact that the equations (\ref{oscal})
and (\ref{ofer}) contain constraints
that to some extend fix behavior of the fields on the global
Cauchy surface. Usual space-time is identified
with the fibration $R^1 \times \gs$ over the local Cauchy surfaces
parametrized by the
time parameter $t$. The Cauchy problem in the generalized
space-time $\M_M$ is defined in terms of a set of functions
on $E$ that allow for localization in terms of distributions
on $\sigma$. The causality requires the propagation in the
 space-time $R\times \gs$ to be microcausal. Remarkably,
the coordinates of the local Cauchy surface compatible with
microcausality
turn out to be associated with the subspace of the symmetric matrices
satisfying the Clifford algebra relations.
This is how the ordinary Minkowski
coordinates and the spinor interpretation of the indices
$\ga, \gb$ (for $M = 2^p$) reappear. Let us note that the
concept of the global Cauchy surface is associated with generalized
space-time ($\M_M$ in our case) while the concept of local  Cauchy
bundle is dependent on a particular dynamical system in $\M_M$. In other
words, different dynamical systems can provide different visualizations
of the same generalized space-time $\M_M$ via different local Cauchy
bundles.

The compactification $\C\M_M= Sp(2M)/P_M$ of $\M_M$ is introduced,
with $P_M$ being an appropriate parabolic subgroup of $Sp(2M)$.
It is shown how the compactified generalized space-time $\C\M_M$
contains infinities of $\M_M$ associated with the singularities of the
generalized inversion in $\M_M$. The formulation of dynamics
in the generalized space-time provides geometric interpretation of
the classical electro-magnetic duality group as the subgroup of the
generalized Lorentz transformations $SL_M\in Sp(2M)$ that leaves invariant
the space-time $R\times \gs$.

For the lower values of $M$, namely $M=2$ and $M=4$, the Lorentz
content of the equations (\ref{oscal}) and (\ref{ofer}) is completely clear.
$\M_2$ identifies with the usual $3d$ space-time $\Sigma= E= \gs = R^2$.
For $\M_4$, $E = \gs \times S^1$ where $\gs = R^3$ is the usual
three-dimensional space while functions on $S^1$ parametrize various
fields of all spins in the usual $4d$ space-time. The duality group
$U(1)$ is the extension of the electro-magnetic duality to all $4d$
higher spins.
The $M=8$ and $M=16$ models describe some $d=6$ and $d=10$ dynamical
systems with $\gs = R^5$ and $\gs= R^9$, respectively. Their Lorentz field
content will be given elsewhere \cite{pr}. Presumably, the $M=8$ theory
provides a higher spin extension of the $6d$ superconformal gravity
theory by Hull \cite{Hull}. The $M=8$  classical duality group
inherited from $\M_M$ is $SO(3)$.

The case of $M=32$ with the generalized conformal symmetry
$Sp(64)$ corresponds to some $d=11$ relativistic theory. To study
its possible relationship with $M$ theory is one of the most exciting
directions for the future investigation.
A related question  is to analyze whether for some $M$ there may
exist different sets of local Cauchy surfaces in the same
model that look like different space-times. Presumably, this
could explain duality of different theories as (non-locally equivalent)
different local realizations of the same model in $\M_M$.

The approach proposed in \cite{BHS} and further developed in
this paper operates in terms of a straightforward generalization of
twistors. As a result, solutions of the field equations decompose into
positive and negative frequency parts  associated with the decomposition
of the solution into the holomorphic and antiholomorphic parts in the
complex coordinates as in (\ref{cbfo}) and (\ref{cffo}). This allows
for a natural quantization prescription in terms of creation and
annihilation operators that depend on the twistor variables $\xi_\ga$
as in (\ref{qpr}) and (\ref{qprf}). The  svector indices $\ga=1\ldots M$
appear in a quite uniform way for all even $M$. Since the svector indices
turn out to be identified with some sets of space-time spinor indices
via the Clifford realization of the Minkowski space-time,
this may have a number of important improvements in the situations in
which the difference between
spinor and tensor representations in the Minkowski track plays a role,
like in dimensional regularization and supersymmetry. The
dynamics in the generalized space admits an extension to generalized
supersymmetric models exhibiting $OSp(L,2M)$ supersymmetries
realized in the appropriate superspaces.
Let us emphasize that only very special Minkowski relativistic
models, like, e.g., the model of all massless $4d$
fields, allow for the realization in the generalized space-time $\M_M$
with unbroken generalized symmetries.

One of the lessons of our analysis is that invariant local field equations
formulated in a space $S$ can effectively describe propagation in
smaller spaces $s$ associated with $S$.
By local observations one can
only observe $s$. However, the full space $S$ manifests itself via
symmetries and specific particle spectra of the theory. In the model
under consideration $S=\M_M$ and $s$ is some Minkowski
space-time. One can say that the usual Minkowski space-time is a
visualization of the generalized space-time $\M_M$. It is tempting
to speculate  that we live in a generalized space-time $\M_M$
which cannot be seen by local observations, but manifests itself via
dualities.

An important question is what is a Lagrangian form of the dynamics in
the generalized space-time. An interesting option
somewhat reminiscent of the group manifold approach \cite{GMA}
is that the Lagrangian is a functional on the submanifold
$R^1 \times\gs$ associated with the usual space-time. To proceed in this
direction it is at any rate  necessary to develop the formulation of the
dynamical equations in terms of potentials rather than in terms of
the fields $b(X)$ and $f_\ga (X)$ which, from the perspective of the
usual space-time, are interpreted as generating functions to the field
strengths (like Maxwell field strength, Weyl tensor  and their further
higher spin generalizations). Introducing the generalized gauge field
(to contain spin one potential, spin 2 metric tensor etc) will
presumably break
down the generalized conformal symmetry transformations $Sp(2M)$
of the equations (\ref{oscal}) and (\ref{ofer}) to a smaller
Poincare or $AdS$-type symmetry.

The generalized $AdS$-like
space-time with $\half M (M+1)$ coordinates
was identified in \cite{BHS} with the group manifold $Sp(M)$.
The $AdS$-type symmetry algebra
associated with the left and right actions of $Sp(M)$ on itself
is $sp(M)\oplus sp(M)$.
Its Lorentz subalgebra  $sp^{l}(M)$ identifies with the diagonal $sp(M)$
while $AdS$ translations belong to the coset space
$sp(M)\oplus sp(M) /sp^{l}(M)$. The conformal symmetries extend
$sp(M)\oplus sp(M)$ to $sp(2M)$.
For $M=2$ one recovers the usual
$3d$ embedding $o(2,2)\sim sp(2)\oplus sp(2) \subset sp(4)\sim o(3,2)$.

The commutation relations of the generalized $AdS$ space-time
symmetries are
\be
\label{LL}
\f{1}{i}[ L_{\ga\gb}\,,L_{\gga\gd} ] =
V_{\gb\gga}L_{\ga\gd}+
V_{\ga\gga}L_{\gb\gd}+
V_{\gb\gd}L_{\ga\gga}+
V_{\ga\gd}L_{\gb\gga}\,,
\ee
\be
\label{LP}
\f{1}{i}[ L_{\ga\gb}\,,P_{\gga\gd} ] =
V_{\gb\gga}P_{\ga\gd}+
V_{\ga\gga}P_{\gb\gd}+
V_{\gb\gd}P_{\ga\gga}+
V_{\ga\gd}P_{\gb\gga}\,,
\ee
\be
\label{PP}
\f{1}{i}[ P_{\ga\gb}\,,P_{\gga\gd} ] = \lambda^2 \Big (
V_{\gb\gga}L_{\ga\gd}+
V_{\ga\gga}L_{\gb\gd}+
V_{\gb\gd}L_{\ga\gga}+
V_{\ga\gd}L_{\gb\gga}\Big )\,,
\ee
where $V_{\ga\gb}= -V_{\gb\ga} $ is a $Sp(M)$ invariant symplectic form
and $\lambda^2$ is a generalized cosmological constant parameter.
Note that, in the generalized space-times, the Lorentz-type subalgebra
$sl_M$ of the conformal algebra is larger than the Lorentz
subalgebra $sp(M)$ of the generalized $AdS$ or Poincare algebras.

The generalized Poincare symmetry results from the limit $\lambda \to 0$
and consists of translations that shift
the coordinates
\be
X^{\ga\gb} \to X^{\prime \ga\gb}  =  X^{\ga\gb} + a^{\ga\gb} \nn
\ee
and $Sp(M)$ ``Lorentz rotations" that leave invariant the
antisymmetric invariant form $V^{\ga\gb}$.
The full list of $\half M$ independent
invariants under the generalized Poincare transformations
consists of
\be
\chi_n = trX^{2n}\,,\qquad n=1\ldots \half M\,,\nn
\ee
where the matrix $X_\ga{}^\gb$ is defined with the aid of the
symplectic form $V_{\ga\gb}$
\be
X_\ga{}^\gb = V_{\gga\ga}X^{\gga\gb}\,. \nn
\ee
(Note that traces of odd powers of $X_\ga{}^\gb$ vanish by antisymmetry
of the symplectic form $V_{\ga\gb}$.) In particular, $\chi_1$ is bilinear in
the coordinates $X$ and  can be identified with the
Lorentz-like interval of the generalized space-time.

A new point compared to the usual Minkowski
geometry is that the metric tensor in the generalized
space-time is not an independent
object being built from the symplectic form $V_{\ga\gb}$
\be
\label{met}
\eta^{\ga\gb ,\gga\gd}
                       = \half \Big ( V^{\ga\gga} V^{\gb\gd}+
V^{\ga\gd} V^{\gb\gga} \Big )\,.
\ee
This metric tensor allows one to single out
generalized Poincare invariant Klein-Gordon and Dirac
equations  from  (\ref{oscal}) and (\ref{ofer})
\be
\label{kg}
\eta^{\ga\gb ,\gga\gd}
\f{\p^2}{\p X^{\ga\gb} \p X^{\gga\gd}} b(X) =0\,,
\ee
\be
\label{dir}
\eta^{\ga\gb ,\gga\gd} \f{\p}{\p X^{\ga\gb}} f_\gga(X)      =0\,.
\ee

The special form of the metric (\ref{met})
may affect the concepts of general relativity
in the generalized space-time:
the corresponding Riemannian geometry is expected to  be very restricted.
Perhaps, this fact may be important for the search of the
$M-$interactions in the conformal theories in higher dimensions \cite{Hull}.
Note that although being based on a symplectic form
$V^{\ga\gb}$, a generalized curved geometry  is expected to have
little to do with the usual symplectic geometry
because the space-time coordinates are symmetric tensors
rather than vectors.

\section*{Acknowledgments}
The author is grateful to A.Barvinsky, V.Fainberg, O.Gelfond,
J.Lukierski, R.Metsaev,  M.Olshanetsky,
V.Ritus, O.Shaynkman, A.Smirnov,
V.Zaikin, all participants of the Monday seminar at INR
and Friday seminar at I.E.Tamm Department of Theoretical
Physics of the Lebedev Institute and especially to V.Rubakov,
M.Soloviev, I.Tyutin and B.Voronov for useful comments and
stimulating discussions.
This research was supported in part by INTAS, Grant No.00-00-254,
the RFBR Grant No.99-02-16207 and the RFBR Grant No.01-02-30024.

\section*{References}
\addcontentsline{toc}{section}{\numberline{}References}

\end{document}